\documentclass[lettersize,journal]{IEEEtran}
\usepackage{amsmath,amsfonts}
\usepackage{algorithmic}
\usepackage{algorithm}
\usepackage{array}
\usepackage[caption=false,font=normalsize,labelfont=sf,textfont=sf]{subfig}
\usepackage{textcomp}
\usepackage{stfloats}
\usepackage{url}
\usepackage{verbatim}
\usepackage{graphicx}
\usepackage{cite}
\usepackage{multirow} 
\usepackage{soul}
\usepackage{xcolor}
\usepackage[inkscapelatex=false]{svg}
\usepackage{rotating}
\usepackage{tabularx}
\usepackage{nicematrix}

\begin{document}

\title{Bridging Accuracy and Explainability in EEG-based Graph Attention Network for Depression Detection}


\author{Soujanya Hazra and Sanjay Ghosh,~\IEEEmembership{Senior Member,~IEEE}
\thanks{This research is supported by the Faculty Start-up Research Grant (FSRG), IIT Kharagpur, awarded to Dr. Sanjay Ghosh.}
\thanks{Soujanya Hazra, and Sanjay Ghosh are with Department of Electrical Engineering, Indian Institute of Technology Kharagpur, WB 721302, India (e-mail: \texttt{sanjay.ghosh@ee.iitkgp.ac.in}).}
}


\markboth{Under-Review}%
{Hazra \MakeLowercase{\textit{et al.}}: Explainable Graph Neural Network for EEG-based Major Depressive Disorder Detection}


\maketitle
\begin{abstract}
%
Depression is a major cause of global mental illness and significantly influences suicide rates. Timely and accurate diagnosis is essential for effective intervention.  Electroencephalography (EEG) provides a non-invasive and accessible method for examining cerebral activity and identifying disease-associated patterns. We propose a novel graph-based deep learning framework, named Edge-gated, axis-mixed Pooling Attention Network (ExPANet), for differentiating major depressive disorder (MDD) patients from healthy controls (HC). EEG recordings undergo preprocessing to eliminate artifacts and are segmented into short periods of activity. We extract 14 features from each segment, which include time, frequency, fractal, and complexity domains. Electrodes are represented as nodes, whereas edges are determined by the phase-locking value (PLV) to represent functional connectivity. The generated brain graphs are examined utilizing an adapted graph attention network. This architecture acquires both localized electrode characteristics and comprehensive functional connectivity patterns. The proposed framework attains superior performance relative to current EEG-based approaches across two different datasets. A fundamental advantage of our methodology is its explainability. We evaluated the significance of features, channels, and edges, in addition to intrinsic attention weights. These studies highlight features, cerebral areas, and connectivity associations that are especially relevant to MDD, many of which correspond with clinical data. Our findings demonstrate a reliable and transparent method for EEG-based screening of MDD, using deep learning with clinically relevant results. The code will be accessible at https://github.com/soujo/ExPANet.
\end{abstract}


\section{Introduction}
\IEEEPARstart{M}{ajor} depressive disorder (MDD) is a prevalent and incapacitating mental health disorder that impacts more than 350 million individuals worldwide, according to the World Health Organization (WHO) \cite{otte2016depression}.  Marked by enduring melancholy, diminished interest in routine tasks, and compromised cognitive abilities, MDD substantially diminishes quality of life and is a significant factor in the increasing global suicide rates.  WHO reports indicate that MDD accounts for about one million fatalities per year \cite{otte2016depression}.  A timely and precise diagnosis is crucial for effective intervention and treatment, potentially averting the disorder's progression and preserving lives.
The conventional clinical diagnosis of MDD mostly relies on systematic interviews, behavioral evaluations, and standardized instruments such as the DSM, BDI, and PHQ \cite{yasin2021mdd}.
Nevertheless, they rely on the proficiency of medical practitioners, leading to subjective diagnostic results. In addition to outward symptoms, people with MDD display aberrant patterns of brain activity \cite{bart2013connectivity}. For instance, patients with MDD demonstrate reduced levels of brain activity in comparison to healthy individuals \cite{yasin2021mdd}. Brain activity is generally assessed via spontaneous electrical activity recording techniques, such as EEG, a non-invasive procedure utilizing electrodes positioned on the scalp \cite{zhao2021plugplay} \cite{sun2016prediction}. EEG analysis indicates significant differences in brain activity between healthy persons and those with MDD, encompassing alterations in energy distribution throughout frequency bands and interactions among functional regions and hemispheres \cite{Thoduparambil2020}. Due to its elevated temporal resolution, affordability, and non-invasive characteristics, EEG has garnered significant interest and has emerged as a promising diagnostic instrument for MDD.

However, the issues of data quality and data volume pose significant challenges for MDD diagnosis using EEG.  A significant level of noise is unavoidable during EEG acquisition, complicating the extraction of distinguishing features from unprocessed EEG data.  Significant noise, including ambient and mechanical disturbances, ocular movements, and cardiac abnormalities, is unavoidable during EEG collecting \cite{wang2024ssgcnet}. Certain studies employ pre-processing techniques, like band filtering and independent component analysis (ICA), to eliminate noise from raw EEG data \cite{SADATSHAHABI2021946}, \cite{xie2022transformer}, \cite{wang2023psg}, \cite{pan2018hand}, \cite{gaurav2021dfa}.  Nevertheless, they depend significantly on human inspection, which is labor-intensive and lacks generalizability, and they are unable to eliminate noise. To address the issue effectively, automated techniques should be suggested to enhance the model's resilience to noise.
Recruiting a substantial number of people to gather adequate and varied data for model training is challenging. The challenges in subject recruitment, privacy concerns, and the substantial expenses associated with data cleansing \cite{ho2023seizure} typically result in a small number of individuals (e.g., 30 \cite{kumar2019eeglstm}, 51 \cite{li2019feature}, and 64 \cite{mumtaz2016dataset}, and the range is between 12 and 213 \cite{yasin2021mdd}) in a MDD diagnosis dataset. The clipped EEG samples from the same subject are intimately interconnected, resulting in a lack of variability in the data. \cite{wang2024sleep} While current EEG-based studies on MDD diagnosis can yield some valid results from the limited data, the tiny sample size significantly impairs model performance. The aforementioned issues contribute to the overfitting issue, particularly in deep learning methodologies, which typically rely on data quality and quantity for optimal performance.  Current deep learning techniques for MDD diagnosis utilizing EEG \cite{Saeedi2020knn}, \cite{LEI2022103370} concentrate on direct training with original datasets, resulting in constrained performance.

In this work, we propose a graph-based deep learning system that models EEG data as brain connection graphs for adaptive and interpretable classification of MDD and HC subjects. Our method derives an extensive variety of features from each EEG channel.  These include time-domain, frequency-band, fractal, and complexity metrics. The collected features constitute node properties in a graph, with nodes symbolizing EEG electrodes and edges delineated by phase-locking value (PLV). The graphs are further examined utilizing a graph-based architecture that captures spatial relationships and non-linear feature interactions. Our framework prioritizes explainability. It determines the most significant features, electrodes, and connection patterns that facilitate classification. This enhances interpretability and develops clinician confidence in the model findings.

The main contributions of this work are: 
\begin{enumerate}
    \item \textit{Multi-domain feature integration:} We utilize 14 EEG features from time, frequency, fractal, and complexity domains as node attributes to capture short-range variability and long-range dynamical characteristics.  
    \item \textit{Architecture:} We design a novel graph-based framework, named as \textbf{E}dge-gated, a\textbf{x}is-mixed \textbf{P}ooling \textbf{A}ttention \textbf{Net}work (\textbf{ExPANet}) for EEG graphs that enables the model to utilize feature heterogeneity and functional connectivity for better classification. 
    \item \textit{Explainability:} Our systematic interpretability process goes beyond accuracy:  
    \begin{enumerate}
        \item \textit{Feature importance:} Determines the most discriminative features for MDD.  
        \item \textit{Channel importance:} Demonstrates electrode-specific relevance maps, highlighting critical brain areas associated with depression. 
        \item \textit{Edge importance:} Highlights functional connectivity associations (via PLV) that vary between groups, offering network-level understanding. 
        \item \textit{Attention analysis:} Utilizes intrinsic graph attention weights to verify learned dependencies and explain subgraph-level mechanisms. 
    \end{enumerate}
    \item \textit{Clinical relevance:} Our framework integrates feature-driven modeling with explainability to provide interpretable EEG signs of MDD, therefore balancing deep learning efficiency with neuroscientific explanation.  
\end{enumerate}

This paper is summarized as follows. Section~\ref{sec:related_work} reviews current methodologies for EEG-based MDD detection and graph neural networks. Section~\ref{sec:methodology} presents the proposed framework, which includes feature extraction, graph construction, the proposed ExPANet architecture, and explainability modules. Section~\ref{sec:expt} outlines the experiments, involved datasets, preprocessing, and results, including overall performance and multi-level interpretability. Section~\ref{sec:conc} concludes our research work.

\section{Related Work}
\label{sec:related_work}

\subsubsection{Machine Learning with Handcrafted Features}
For at least the previous two decades, machine learning methodologies have been applied to resting-state EEG data \cite{reza2009sz}. 
Preliminary deep learning-based methodologies for major depressive disorder diagnosis rely on manually created features, such as multi-layer perceptrons (MLP) \cite{DING2019156} and probabilistic neural networks (PNN) \cite{Mahato2019eeg}.
Diverse signal processing techniques \cite{Christopher1998wavelet}, \cite{Jeong2016coherence} are consistently implemented to analyze EEG signals for the understanding of neurological disorders, including Alzheimer’s disease \cite{Ziad2012az} and schizophrenia \cite{ghosh2023connectome}.
These approaches established the foundations for interpretable graph-based models. In epilepsy research, interpretable GNNs have shown the ability to precisely identify seizures while emphasizing pathologically relevant brain connections \cite{Mazurek2024graphxai}. 
\subsubsection{Deep Learning with CNN and Hybrid Architectures}
Recent years have seen various studies focusing on the diagnosis of MDD through EEG, applying deep learning methodologies. Through the analysis of EEG signals, researchers can investigate localized brain activity and patterns of functional connectivity. This indicates network-level interactions, offering insights into potential features unique to MDD \cite{Friston2011connectivity}, \cite{ellis2024crosseeg}.
Recent models implementing CNN architectures entirely extract characteristics defined by local invariance, especially in the short term. In this regard, DeprNet \cite{seal2021deprnet} employs the ConvNet architecture, with five stacked CNN layers, to extract temporal features.
InceptionNet \cite{LEI2022103370}, \cite{Uyulan2021cnn} utilizes kernels of diverse dimensions to extract features across many scales and implements channel-wise attention in the higher levels to determine channel importance.
In this study \cite{SADATSHAHABI2021946}, the authors first extract spectral features and then integrate the time-frequency maps into a 2D-CNN.  
Recently, because of LSTM's capacity to store long-term characteristics, certain studies have utilized hybrid architectures. 
The CNN-LSTM architecture, which is used to represent both short-term and long-term characteristics at the same time, is developed in \cite{Saeedi2020knn} and \cite{sharma2021dephnn}.
In order to generate brain connection graphs in an adaptive manner, the authors \cite{luo2023graph} incorporate the GCN and GRU models. This integration allows it to extract the spatial-temporal properties of the EEG. 
Current research trains models using original data directly, and their effectiveness is constrained by inadequate data quality and insufficient data volume in the diagnosis of MDD utilizing EEG. 
Recent developments in computational neuroscience have provided new insights into how underlying anatomy may limit EEG-derived functional connectivity.   While concurrently attempting to bring together the anatomical and functional brain connectomes \cite{ghosh2023subspace}.
In this paper \cite{lawhern2018eegnet}, the authors introduced a deep learning architecture for EEG-based identification, named EEGNet. Despite significant advancements, the majority of prior methodologies either concentrate on temporal/spectral patterns or utilize high-dimensional raw representations that lack interpretability. 

Limited research has employed graph-based frameworks for the diagnosis of MDD to dynamically model the spatial, spectral, and nonlinear dynamics of EEG signals.  
Recent research on explainability sometimes lacks clinical evidence or fails to establish the significance of certain brain regions and connections.  
Our present work incorporates EEG-derived feature graphs and interpretable graph learning to enhance the diagnostic performance and clinical relevance of MDD classification.
\begin{figure*}[htbp]
    \centering 
    \includegraphics[width=1\textwidth]{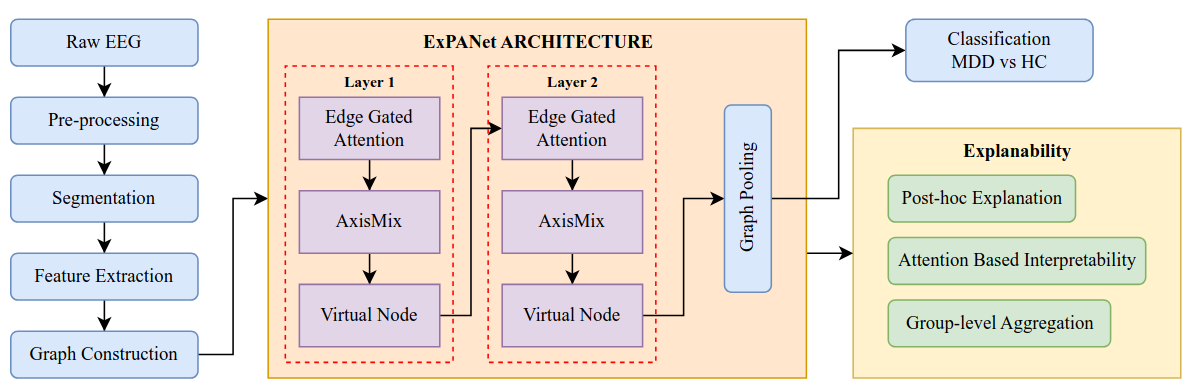}
    \includegraphics[width=1\textwidth]{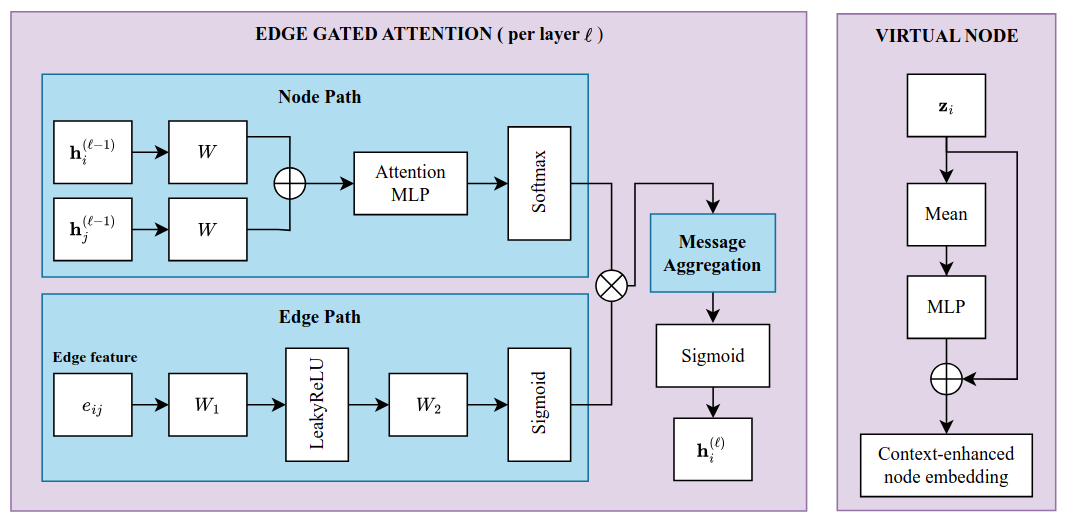}
    \includegraphics[width=1\textwidth]{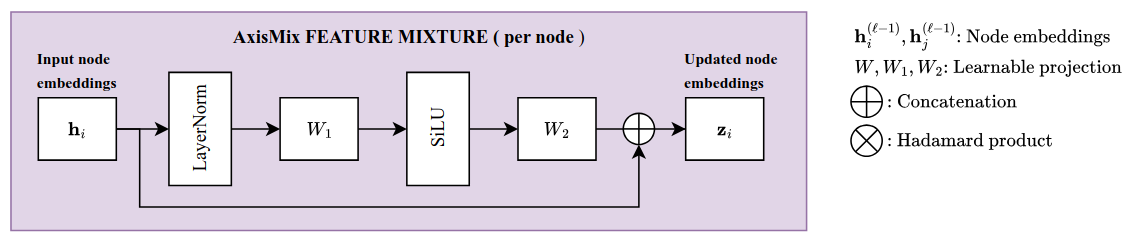}
    \caption{Illustration of the proposed ExPANet architecture for EEG-based classification of MDD vs HC. 
    }
    \label{fig:expanet}
\end{figure*}

\section{Methodology}
\label{sec:methodology}
%
Our framework is designed to address three fundamental components of EEG-based mental diagnosis: artifact elimination, functional connectivity modeling, and interpretability. The mechanism operates in multiple phases. Initially, EEG signals undergo preprocessing to eliminate noise and artifacts. After the recordings, they are divided into epochs of defined duration. We extract 14 features from each segment, including the time, frequency, fractal, and complexity domains. These representations constitute the node properties of EEG electrode graphs. Edges connecting nodes are characterized by the PLV, which quantifies phase synchronization among channels. This generates connectivity graphs that are particular to subjects and segments, illustrating across-region communication within the brain. The graphs are examined utilizing our adapted graph attention network, ExPANet. This approach integrates local feature data with global network architecture, facilitating precise categorization of MDD vs HC. The architecture integrates attention techniques and edge gating to improve learning from node attributes and connections. This framework offers interpretability at various levels, in addition to classification. We determine the most significant features, electrodes, and functional linkages for predictive analysis.  This enables us to map decision pathways and correlate them with established neurophysiological results in depression.
Fig.\ref{fig:expanet} provides a visual representation of the primary computational flow and internal modules that constitute the ExPANet framework.
This approach is ideal for practical and explainable EEG-based diagnosis of MDD due to its robust classification accuracy and clinically relevant explanations.
All mathematical symbols and variables used in this paper are defined in Table~\ref{tab:notation}.
\begin{table}[htbp]
\centering
\caption{Symbols and notations used throughout the paper.}
\label{tab:notation}
{\renewcommand{\arraystretch}{1.4}%
\begin{tabular}{|m{0.3cm}|m{2.4cm}|m{4.7cm}|}
\hline
\textbf{Gr.} & \textbf{Symbol} & \textbf{Meaning}\\
\hline

\multirow{6}{*}{\rotatebox[origin=c]{90}{\textbf{Graph}}} 
& $\mathcal{G}=(\mathcal{V},\mathcal{E},\mathbf{X})$
& EEG brain graph with nodes $\mathcal{V}$, edges $\mathcal{E}$, and feature matrix $\mathbf{X}\in\mathbb{R}^{|\mathcal{V}|\times F}$ \\
\cline{2-3}
& $\mathbf{A}$
& Weighted adjacency matrix $[0,1]^{|\mathcal{V}|\times|\mathcal{V}|}$ derived from PLV \\
\cline{2-3}
& $\mathbf{h}_i^{(\ell)}$
& Node embedding of node $i$ at layer $\ell$ ($d_\ell$) \\
\cline{2-3}
& $\mathcal{N}(i)$
& Neighbor set of node $i$ in graph $\mathcal{G}$ \\
\cline{2-3}
& $L$
& Total number of GNN layers \\
\cline{2-3}
& $F$
& Number of EEG features per node (14) \\
\hline

\multirow{6}{*}{\rotatebox[origin=c]{90}{\textbf{Connectivity}}}
& $\phi_i(t)$
& Instantaneous phase of EEG channel $i$ at time $t$ \\
\cline{2-3}
& $\mathrm{PLV}_{ij}$
& Phase locking value between channels $i,j$ $\in[0,1]$ \\
\cline{2-3}
& $N_t$
& Number of time samples in each EEG segment \\
\cline{2-3}
& $t_k$ 
& \( k^{\text{th}} \) time point in the sampled segment\\
\cline{2-3}
& $e_{ij}$
& Edge feature derived from PLV between channels $i$ and $j$ \\
\cline{2-3}
& $\mathcal{E}$
& Set of all undirected edge pairs $(i,j)$ \\
\hline

\multirow{8}{*}{\rotatebox[origin=c]{90}{\textbf{Attention}}}
& $W^{(\ell)}$
& Linear projection matrix at layer $\ell$ ($d_\ell\times d_{\ell-1}$) \\
\cline{2-3}
& $\mathbf{a}$
& Learnable attention vector ($2d_\ell+1$) \\
\cline{2-3}
& $\alpha_{ij}^{(\ell)}$
& Normalized attention coefficient between nodes $i,j$ at layer $\ell$ \\
\cline{2-3}
& $\phi(\cdot)$
& LeakyReLU activation used in attention scoring \\
\cline{2-3}
& $g_{ij}$
& Edge-gate coefficient $\in(0,1)$ \\
\cline{2-3}
& $\mathbf{v}, \mathbf{v}'$
& Virtual node state and its updated representation ($d_\ell$) \\
\cline{2-3}
& $\mathrm{LN}(\cdot)$
& Layer Normalization operation \\
\cline{2-3}
& $z_i$
& AxisMix-modified node embedding ($d_\ell$) \\
\hline

\multirow{5}{*}{\rotatebox[origin=c]{90}{\textbf{Pooling}}}
& $\mathbf{g}_{\text{mean}}$
& Mean-pooled graph vector ($d_L$) \\
\cline{2-3}
& $\mathbf{g}_{\text{add}}$
& Sum-pooled graph vector ($d_L$) \\
\cline{2-3}
& $\mathbf{g}_v$
& Virtual-node embedding ($d_L$) \\
\cline{2-3}
& $\mathbf{g}$
& Final concatenated graph embedding ($3d_L$) \\
\cline{2-3}
& $\hat{y}_m$
& Predicted probability for sample $m$ \\
\hline

\multirow{4}{*}{\rotatebox[origin=c]{90}{\textbf{Loss}}}
& $y_m$
& Ground-truth label $\in\{0,1\}$ for sample $m$ \\
\cline{2-3}
& $M$
& Mini-batch size \\
\cline{2-3}
& $\sigma(\cdot)$
& Sigmoid activation function \\
\cline{2-3}
& $\mathcal{L}$
& Binary cross-entropy classification loss \\
\hline

\multirow{9}{*}{\rotatebox[origin=c]{90}{\textbf{Explainability}}}
& $M_E,M_V,M_F$
& Learnable edge, node, and feature masks \\
\cline{2-3}
& $\pi_E,\pi_V,\pi_F$
& Sigmoid-relaxed masks in $(0,1)$ \\
\cline{2-3}
& $\tilde{\mathbf{A}}$
& Masked adjacency \\
\cline{2-3}
& $\tilde{\mathbf{X}}$
& Masked feature matrix \\
\cline{2-3}
& $\mathcal{L}_{\text{fid}}$
& Fidelity loss ensuring prediction consistency \\
\cline{2-3}
& $\mathcal{L}_{\text{reg}}$
& Regularization loss  \\
\cline{2-3}
& $\mathcal{D}_g$
& Graph set for group $g\!\in\!\{\mathrm{HC},\mathrm{MDD}\}$ \\
\cline{2-3}
& $\bar{m}_F^{(g)}, \bar{m}_V^{(g)},  \bar{m}_{E,(i,j)}^{(g)}$
& Group-averaged feature, node, and edge saliencies \\
\cline{2-3}
& $f_\theta$
& Trained GNN model with learnable parameters $\theta$ \\
\hline

\end{tabular}}
\end{table}
\subsection{EEG Feature Extraction}
We extract a variety of EEG features from each 5-second preprocessed EEG segment to assess MDD brain processes. These features measure signal properties in time, frequency, fractal, and complexity domains. EEG channels are processed separately, yielding a total of 14 features per channel, resulting in a \(19\times 14\) feature matrix per segment. These qualities become brain graph node attributes. A compact and interpretable representation of local signal changes and non-linear brain dynamics is achieved using this multi-domain technique.

Let \( x = \{x_1, x_2, \ldots, x_N\} \) represent the EEG signal from one channel over a 5-second segment.
\subsubsection{Variance}
Variance is a crucial statistical measure that assesses local brain activity within a specified time interval. 
The variance of the EEG signal is computed as:
\begin{equation}
\text{Var}(x) = \frac{1}{N} \sum_{i=1}^{N} (x_i - \bar{x})^2,
\label{eq:variance}
\end{equation}
where \( \bar{x} \) is the mean amplitude of the signal and \( N \) is the number of samples. Clinically, frontal electrode variation may indicate hypoactivity or abnormal cortical involvement \cite{xiaowei2016feature} in MDD patients, helping to distinguish them from HC.

\subsubsection{Line length}
Line length is a basic yet effective time-domain descriptor that sums the absolute differences across samples to describe EEG signal variation as follows:  
\begin{equation}
\text{LL}(x) = \sum_{i=1}^{N-1} |x_{i+1} - x_i|,
\label{eq:linelength}
\end{equation}
where \( N \) is the number of samples. Changes in EEG complexity, reflected through line length \cite{Esteller2001linelength}, may indicate alterations in neural synchrony and variability patterns often reported in mood disorders like depression.
\subsubsection{Hjorth parameters}
Time-domain Hjorth parameters \cite{Hjorth1970prop} quantify signal frequency and complexity without spectral transformation. 
Hjorth mobility characterizes the signal's mean frequency content as follows:
\begin{equation}
\text{Mobility}(x) = \sqrt{ \frac{ \text{Var}(\dot{x}) }{ \text{Var}(x) } },
\label{eq:hjorth_mobility}
\end{equation}
where \( \dot{x} \) denotes the first derivative of the EEG signal \( x \). Hjorth complexity (temporal variations in a signal's frequency) is:
\begin{equation}
\text{Complexity}(x) = \frac{ \text{Mobility}(\dot{x}) }{ \text{Mobility}(x) },
\label{eq:hjorth_complexity}
\end{equation}
These factors represent the dynamic properties of EEG signals. 
\subsubsection{Katz Fractal Dimension (KFD)}
In the EEG signal, KFD denotes the degree of irregularity or "chaotic" nature of voltage variations across time within a particular brain region. 
The KFD is calculated as follows:
\begin{equation}
\text{KFD}(x) = \frac{ \log_{10}(L/a) }{ \log_{10}(d/a) },
\label{eq:kfd}
\end{equation}
where \( L = \sum_{i=1}^{N-1} |x_{i+1} - x_i| \) is the total waveform length; \( d = \max |x_i - x_1| \) is the maximum distance from the first sample point; and  \( a = L / (N - 1) \) is the average step length between the points.
%
%
We calculate KFD independently for each of the 19 EEG channels, resulting in one scalar per channel for the feature vector. 
Esteller et al. \cite{Esteller2001fractal} shown that KFD serves as a sensitive marker for aberrant brain states, and its resilience in noisy environments renders it a viable feature for automated depression screening.

\subsubsection{Higuchi Fractal Dimension (HFD)}
HFD is considered a sensitive feature for assessing irregularities in brain activity.
Clinically, lower HFD scores may reflect cognitive flexibility, emotional regulation, or information processing issues, typical of depression \cite{Higuchi1988fractal}.  
To analyze temporal scaling behavior, the approach \cite{Higuchi1988fractal} generates downsampled subseries at different temporal resolutions and evaluates their average geometric length.  The resultant fractal dimension shows the growth of EEG wave-form complexity, revealing the brain's adaptive neural responses.
\subsubsection{Detrended Fluctuation Analysis (DFA)}
We apply a scaling analysis method to identify long-range temporal correlations in non-stationary EEG signals.  We measure the variation of the root-mean-square fluctuation of the integrated and detrended signal across various time scales.  In the setting of depression, modified DFA exponents correlate with impaired neural autocorrelations and diminished signal complexity, especially in the frontal and temporal areas of the brain \cite{Bryce2012dfa}.


\subsubsection{Lempel-Ziv Complexity (LZC)}
In EEG analysis, LZC \cite{Lempel1976ziv} quantifies the number of unique patterns in a binary-transformed time series, indicating the signal's level of regularity or compressibility. The EEG time series is initially binarized and then the binary string is analyzed to determine the quantity of distinct subsequences observed during a left-to-right examination. This metric quantifies temporal anomalies in the cerebral signal at the symbolic level.
%
%
Reduced LZC values are commonly noted in pathological states such as MDD, indicating a decline in brain dynamical diversity and information transmission capacity \cite{ADERINWALE2023EEG}.
\subsubsection{Permutation Entropy (PE)}
It is a nonlinear complexity metric that measures the temporal arrangement of amplitude values in a time series. In contrast to conventional entropy-based metrics,
it resilient to noise and particularly appropriate for brief, non-stationary physiological data such as EEG \cite{Bandt2002PE}. 

PE is computed as a normalized Shannon entropy measure:
\begin{equation}
    \text{PE} = -\frac{1}{\log(d!)} \sum_{i=1}^{d!} p_i \log(p_i)
\label{eq:entropy}
\end{equation}
where \( p_i \) is the probability of observing the \( i^{\text{th}} \) permutation among \( d! \) possible patterns. A lower PE indicates more deterministic behavior, while a value closer to 1 reflects higher complexity.
\subsubsection{Band Power (delta, theta, alpha, beta, gamma)}
Band power characteristics measure the energy present in discrete frequency bands of the EEG data, indicating distinct brain oscillations linked to various cognitive and emotional functions. The five basic frequency bands: delta (0.5–4 Hz), theta (4–8 Hz), alpha (8–13 Hz), beta (13–22 Hz), and gamma (22–30 Hz) are frequently utilized as spectral biomarkers in psychiatric and neurological studies \cite{Klimesch1999cognitive}. In the setting of MDD, anomalies in band-specific energy, particularly elevated theta and diminished alpha or beta power, have been observed in fronto-central areas. These aberrations are associated with impaired circuitry for arousal, attention, and affective control in the brain.
To calculate band power, a \(4^{\text{th}}\) order Butterworth bandpass filter is utilized for each frequency range, and the logarithmic energy of the filtered signal is determined as:
\begin{equation}
    E_{b} = \log \left( \sum_{t=1}^{N} x_b^2(t) \right)
\label{eq:energy}
\end{equation}

where \( x_b(t) \) is the bandpass-filtered EEG signal in band \( b \), and \( N \) is the number of time samples. The logarithmic transformation helps stabilize variance and reduce skewness across subjects.

\subsection{Graph Construction from EEG Features}
\label{subsec:graph_cons}
To acquire intricate representations of EEG signals for efficient classification of MDD, we attempt to explore graph neural network (GNN) learning. We construct an undirected weighted graph \( \mathcal{G} = (\mathcal{V}, \mathcal{E}, \mathbf{X}) \), where each node \( \mathcal{V} = \{1, 2, \dots, N\} \) represents the EEG channel and \( \mathcal{E} \subseteq \mathcal{V} \times \mathcal{V} \) denotes the set of edges representing the functional correlation of the EEG time-series (5-seconds long) between each pair of channels. 
%
Each node \( i \in \mathcal{V} \) is associated with a feature vector \( \mathbf{h}_i^{(0)} \in \mathbb{R}^F \), where \( F = 14 \) denotes the number of features extracted from the corresponding EEG signal. These features span multiple signal domains and include:
(i) \textit{Time-domain descriptors:} variance, line length, Hjorth parameters, DFA, and LZC; (ii) \textit{Frequency-domain metrics:} bandpower in delta, theta, alpha, beta, and gamma bands; (iii) \textit{Non-linear/fractal measures:} KFD, HFD, and entropy.
%
The feature vectors are stacked to form the node attribute matrix \( \mathbf{X}^{(0)} \in \mathbb{R}^{N \times F} \), where each row corresponds to one EEG channel. 

Graph edges $\mathcal{E}$ are constructed using phase locking value (PLV) \cite{Ricardo2018PLV}, a recognized metric for phase synchronization between two brain signals.
%
Let \( \phi_i(t) \) and \( \phi_j(t) \) denote the instantaneous phases of channels \( i \) and \( j \) at time \( t \), obtained via the Hilbert transform.
The PLV value between nodes \( i \) and \( j \) is defined as:
%
%
\begin{equation}
    \text{PLV}_{ij} = \left| \frac{1}{N_t} \sum_{k=1}^{N_t} e^{j(\phi_i(t_k) - \phi_j(t_k))} \right|
\label{eq:plv}
\end{equation}
where $N_t$ is the number of time samples in the segment and $t_k$ denotes the \( k^{\text{th}} \) time point in the sampled segment.
%
For each EEG segment, the PLV value which is in the range $[0, 1]$, is calculated pairwise for all distinct electrode pairings. Finally, the symmetric adjacency matrix \( \mathbf{A} \in \mathbb{R}^{19 \times 19}\).
%
To minimize unnecessary connections and ensure graph sparsity, we perform thresholding by retaining only the top-\(k\) PLV edges for each node.

\subsection{Proposed Graph Neural Network Architecture}  

To diagnose major depressive disorder (MDD) using EEG data, we propose a novel framework, named as, \textbf{E}dge-gated, a\textbf{x}is-mixed \textbf{P}ooling \textbf{A}ttention \textbf{Net}work (\textbf{ExPANet}).
Our architecture presents three innovations: (i) a \emph{edge-gated attention} mechanism for optimizing connectivity strengths, (ii) a \emph{AxisMix} module for feature-axis integration, and (iii) a \emph{virtual node} for the incorporation of global context. A triple pooling technique integrates various graph-level representations for enhanced classification robustness. This subsection presents the detailed information on the architecture.

\subsubsection{Edge-Gated Attention Propagation}
Suppose that at layer $(\ell-1)$, the embedding of node $i$ is $\mathbf{h}_i^{(\ell-1)} \in \mathbb{R}^{d_{\ell-1}}$. Now in the next layer $\ell$, the embedding is updated as:
\begin{equation}
\mathbf{h}_i^{(\ell)} = \sigma \!\left( \sum_{j \in \mathcal{N}(i)} g_{ij}\,\alpha_{ij}^{(\ell)} W^{(\ell)} \mathbf{h}_j^{(\ell-1)} \right),
\label{eq:edgegatupdate}
\end{equation}
where $W^{(\ell)} \in \mathbb{R}^{d_\ell \times d_{\ell-1}}$ is a learnable projection, 
$\sigma(\cdot)$ is a nonlinearity, and $g_{ij} \in (0,1)$ is a learned edge gate.  
%
The attention coefficient $\alpha_{ij}^{(\ell)}$ augmented with edge attributes $e_{ij}$ is computed as
\begin{equation}
\alpha_{ij}^{(\ell)} =
\frac{
\exp\!\left(
\phi\!\left(
\mathbf{a}^{\top}
\!\left[
W^{(\ell)}\mathbf{h}_i^{(\ell-1)}
\ | \
W^{(\ell)}\mathbf{h}_j^{(\ell-1)}
\ | \
e_{ij}
\right]
\right)
\right)
}{
\sum\limits_{k \in \mathcal{N}(i)}
\exp\!\left(
\phi\!\left(
\mathbf{a}^{\top}
\!\left[
W^{(\ell)}\mathbf{h}_i^{(\ell-1)}
\ | \
W^{(\ell)}\mathbf{h}_k^{(\ell-1)}
\ | \
e_{ik}
\right]
\right)
\right)
},
\label{eq:edgeattention}
\end{equation}
where $\mathbf{a}\!\in\!\mathbb{R}^{2d_\ell+1}$ is a learnable attention vector, 
$\phi(\cdot)$ denotes the LeakyReLU nonlinearity, $e_{ij}$ is the PLV-derived edge feature and $[\,\cdot\, |\,\cdot\,]$ represents the vector concatenation. Notice that normalization is performed on all neighbors $k\!\in\!\mathcal{N}(i)$ of node $i$.
The gate $g_{ij}$ is generated from $e_{ij}$ by:
\begin{equation}
g_{ij} = \sigma \!\big( W_2 \,\phi(W_1 e_{ij}) \big),
\end{equation}
with $W_1 \in \mathbb{R}^{h \times 1}$ and $W_2 \in \mathbb{R}^{1 \times h}$.  
Using this formulation, we generalize edge-aware GNNs and thus allow dynamic control of functional connectivity.

\subsubsection{AxisMix Feature Mixer}
After attention aggregation, node embeddings are passed through AxisMix as follows:
\begin{equation}
\mathbf{z}_i = \mathbf{h}_i^{(\ell)} + W_2\,\phi(W_1 \,\text{LN}(\mathbf{h}_i^{(\ell)})),
\end{equation}
where LN is layer normalization, 
$W_1 \in \mathbb{R}^{d_\ell \times (2d_\ell)}$, 
$W_2 \in \mathbb{R}^{(2d_\ell) \times d_\ell}$, 
and $\phi(\cdot)$ is SiLU activation. 
Our AxisMix module captures higher-order interactions among feature characteristics, while residual linkages maintain stability. 

\subsubsection{Virtual Node Mechanism}
To integrate global context into each ExPANet layer, we develop a virtual node that combines information from all current node embeddings and redistributes it to each node.
Given the locally updated embeddings $\{\mathbf{z}_i\}_{i=1}^N$ obtained after the AxisMix block, 
the global summary vector $\mathbf{v}$ is computed as the mean across all nodes:
\begin{equation}
\mathbf{v} = \frac{1}{N}\sum_{i=1}^N \mathbf{z}_i.
\end{equation}
This global representation is then refined by a small multilayer perceptron:
\begin{equation}
\mathbf{v}' = \phi(W_v \mathbf{v} + b_v),
\end{equation}
where $\phi(\cdot)$ denotes a nonlinear activation. 
The updated global context $\mathbf{v}'$ is distributed and incorporated into each node embedding via a residual update:
\begin{equation}
\mathbf{h}_i^{(\ell)} = \mathbf{z}_i + \mathbf{v}', \quad \forall i \in \{1, \ldots, N\}.
\end{equation}
This virtual node approach allows all nodes to share information, representing long-range dependency between spatially distant EEG electrodes and improving overall brain dynamics.

\subsubsection{Graph-level Pooling and Classification}
Following two iterations of edge-gated attention and AxisMix, and then a virtual node update, we perform graph embedding. Suppose $L$ denotes the total number of graph layers.
We integrate three pooling strategies:
\begin{align}
\mathbf{g}_{\text{mean}} = \frac{1}{N}\sum_{i=1}^N \mathbf{h}_i^{(L)}, \quad 
\mathbf{g}_{\text{add}}  = \sum_{i=1}^N \mathbf{h}_i^{(L)}, \quad 
\mathbf{g}_v = \mathbf{v}'.
\end{align}
The final representation is given by:
\begin{equation}
\mathbf{g} = [\mathbf{g}_{\text{mean}} \ | \ \mathbf{g}_{\text{add}} \ | \ \mathbf{g}_v] \in \mathbb{R}^{3d_L}. 
\end{equation}
Finally, this is fed into an MLP as follows:
\begin{equation}
\hat{y} = \sigma \big( W_3 \,\phi \big(W_2 \,\phi(W_1 \mathbf{g} + b_1)+b_2 \big) + b_3 \big),
\end{equation}
where $\hat{y}$ represents the likelihood of classifying as MDD, 
and $\sigma(\cdot)$ is the sigmoid activation function. We employ binary cross-entropy loss:
\begin{equation}
\mathcal{L} = -\frac{1}{M} \sum_{m=1}^{M} \Big[ y_m \cdot \log \hat{y}_m \;+\; (1 - y_m) \cdot \log \big(1 - \hat{y}_m\big) \Big],
\end{equation}
where $M$ is the batch size, $y_m \in \{0,1\}$ is the ground-truth label for the $m^{\text{th}}$ sample, and 
$\hat{y}_m \in \mathbb{R}$ is the raw logit prediction probability by our proposed deep network.

Our architecture combines local channel dynamics and long-range synchronization. Edge-awareness improves channel connection. AxisMix enhances feature interactions. Each segment receives global context from the virtual node. The techniques capture localized EEG activity and distributed functional connections. This representation is effective in identifying MDD patients from healthy controls, where brain networks change gradually but widely. This balanced performance is essential in medical applications.

\subsection{Explainability in Graph Neural Networks}
Deep learning models for EEG-based psychiatric diagnosis frequently attain high predictive accuracy, but at the expense of transparency.  In clinical applications, interpretability is essential. Physicians must check that the model's decision-making process corresponds with neurophysiological information.  Graph Neural Networks (GNNs) are frequently regarded as "black boxes" due to the distribution of node embeddings and edge transformations over several non-linear layers \cite{ying2019gnnexplainer}.  To resolve this, we integrated post-hoc explainability modules and intrinsic attention-based interpretability into our system.

\subsubsection{Post-hoc Explanation}
%
We utilize GNNExplainer to determine a brief subgraph and feature subset that maintains the model's prediction \cite{ying2019gnnexplainer}. Design a trained GNN as $f_{\theta}$. We know that $\mathcal{G} = (\mathcal{V}, \mathcal{E}, \mathbf{X})$ denotes the input graph (refer to Section~\ref{subsec:graph_cons}), where the adjacency matrix $\mathbf{A} \in [0,1]^{|\mathcal{V}| \times |\mathcal{V}|}$ and the node features matrix $\mathbf{X} \in \mathbb{R}^{|\mathcal{V}| \times F}$. In the context of graph-level classification, establish the reference label as $\hat{y}=\arg\max_{c} f_{\theta}(\mathcal{G})_c$. We produce continuous masks across edges, nodes, and features:
\[
M_E\in\mathbb{R}^{|\mathcal{E}|},\quad 
M_V\in\mathbb{R}^{|\mathcal{V}|},\quad 
M_F\in\mathbb{R}^{F}.
\]
Their sigmoid relaxations are $\pi_E=\sigma(M_E)$, $\pi_V=\sigma(M_V)$, and $\pi_F=\sigma(M_F)$ with entries in $(0,1)$.

\paragraph{Masked graph}
We form a soft-masked graph $\mathcal{G}\odot (M_E,M_V,M_F)$. For each edge $e_k = (i,j) \in \mathcal{E}$, define \( \tilde{\mathbf{A}}_{ij}=\tilde{\mathbf{A}}_{ji}= \pi_{E,k} \mathbf{A}_{ij} \).
For each node $i \in \mathcal{V}$, adjust its features using node and feature masks:
 \[ \tilde{\mathbf{X}}_{i,:}= \pi_{V,i}\,\big(\mathbf{X}_{i,:}\odot \pi_F^{\top}\big). \]
Accordingly, the explanation retains edges and features with substantial mask values while diminishing the rest of the features.

\paragraph{Fidelity objective}
We estimate the mutual-information objective using a prediction-fidelity loss.
 \begin{equation} 
 \label{eq:fid} \mathcal{L}_{\text{fid}} \;=\; -\log p_{\theta}\!\big(\hat{y}\,\big|\, \mathcal{G}\odot(M_E,M_V,M_F)\big). \end{equation}
This justifies $f_{\theta}$ to yield an identical decision on the masked graph as on $\mathcal{G}$.

\paragraph{Sparsity and discreteness}
We standardize mask size and restrict masks towards $\{0,1\}$: \begin{equation} 
\label{eq:reg} 
\begin{split}
\mathcal{L}_{\text{reg}} \;=\; \alpha\,\|\pi_E\|_{1} \;+\; \beta\,\bar{H}(\pi_E) \;+\; \gamma\,\|\pi_F\|_{1} \;+\; \delta\,\bar{H}(\pi_F) \\
\;+\; \eta\,\|\pi_V\|_{1} \;+\; \zeta\,\bar{H}(\pi_V). 
\end{split}
\end{equation}
Here, $\|\cdot\|_1$ denotes the $L_1$ norm.
The expression $\bar{H}(\pi)=\frac{1}{n}\sum_{i} \big[-\pi_i\log\pi_i-(1-\pi_i)\log(1-\pi_i)\big]$ denotes the average entropy of the probability distribution representing the average Bernoulli entropy.
Scalars $\alpha,\gamma,\eta>0$ regulate the sparsity of edges, features, and nodes, respectively.
Scalars $\beta, \delta, \zeta > 0$ regulate binarization.

\paragraph{Final objective and optimization}
The explainer minimizes
\begin{equation}
\label{eq:final}
\min_{M_E,M_V,M_F} \;\; \big( \mathcal{L}_{\text{fid}} \;+\; \mathcal{L}_{\text{reg}} \big).
\end{equation}
The model parameters $\theta$ are fixed. Masks are refined using gradient descent methods, such as Adam, using sigmoid relaxation. This produces almost binary selections of edges, nodes, and features.

\subsubsection{Attention-based Interpretability}
In contrast to post-hoc techniques, attention-based GNNs offer inherent interpretability. In our ExPANet architecture, the edge-gated attention mechanism inherently generates normalized coefficients (refer to \eqref{eq:edgeattention}). This assesses the relative significance of adjacent nodes and their PLV-derived edge properties during message transmission. These coefficients can be readily represented as connection saliency maps, emphasizing functionally important brain regions and inter-channel interactions. Through the comparison of averaged attention maps between the healthy control (HC) and major depressive disorder (MDD) groups, we detect clinically significant disturbances in long-range synchronization and modified frontal–parietal connection.
\subsubsection{Group-level Aggregation}
Individual explanations could demonstrate variation at the segment level. To improve robustness, we merged saliency masks across all graphs within each group (HC or MDD). Let $\mathcal{D}_g$ be the collection of graphs given group $g \in \{\text{HC}, \text{MDD}\}$. The vector of significance at the group level is:
\begin{equation}
\bar{m}_F^{(g)} = \frac{1}{|\mathcal{D}_g|}\sum_{d \in \mathcal{D}_g} \pi_F^{(d)},
\end{equation}
where $\pi_F^{(d)}$ is the feature mask of graph $d$. 
For channels,
\begin{equation}
\bar{m}_V^{(g)} = \frac{1}{|\mathcal{D}_g|}\sum_{d \in \mathcal{D}_g} \pi_V^{(d)}.
\end{equation}
For undirected edges, we construct masks for each graph and subsequently compute the average:
\begin{equation}
\bar{m}_{E,(i,j)}^{(g)} = \frac{1}{|\mathcal{D}_g|}\sum_{d \in \mathcal{D}_g}\max\!\big(\pi_{E,(i \to j)}^{(d)},\,\pi_{E,(j \to i)}^{(d)}\big).
\end{equation}
The combined profiles produce sorted feature lists, channel saliency maps, and circular edge-connectivity diagrams. This multilevel system links raw EEG features with clinically significant group-level anomalies. It delivers consistent interpretations by eliminating individual-level noise and highlighting ongoing group-specific anomalies.
Collectively, these methodologies guarantee that our diagnostic framework is both precise and clinically interpretable, fulfilling the criteria for explainable artificial intelligence in mental healthcare  \cite{Mazurek2024graphxai}.

\section{Experiments}
\label{sec:expt}

\subsection{Datasets}
\label{sec:dataset}

\subsubsection{Dataset-I}
The first dataset is a publicly available EEG dataset \cite{mumtaz2016dataset} recorded at Hospital Universiti Sains Malaysia (HUSM). The dataset contains resting-state EEG recordings from 64 persons, including 34 individuals diagnosed with major depressive disorder (MDD) and 30 healthy controls (HC). The MDD group included 17 males and 17 females (mean age: $40.3 \pm 12.9$ years), whereas the HC group comprised 21 males and 9 females (mean age: $38.3 \pm 15.6$ years). MDD diagnoses were determined according to DSM-IV criteria, and all study protocols received approval from the HUSM ethics committee. Recordings were obtained under both eyes open (EO) and eyes closed (EC) circumstances, with only EC data utilized for subsequent analysis. A total of 28 participants with major depressive disorder (MDD) and 28 healthy controls (HC) were selected based on the duration and quality of recordings.

\subsubsection{Dataset-II}
The second dataset was obtained from the University of New Mexico \cite{cavanagh2019multiple}. Participants were selected from a beck depression inventory (BDI) survey conducted among college-age participants at Arizona State University. The dataset initially comprised 122 patients, of whom 46 were categorized as depressive or displayed high BDI scores. To enhance balance and quality, we selected 23 present or former MDD patients and 23 healthy controls with the lowest BDI scores as target participants. This yielded a total of 46 participants, evenly divided between the major
depressive disorder (MDD) and healthy controls (HC) groups.

\subsection{Pre-processing}
EEG recordings are prone to several anomalies, including ocular movements, muscular activity, and ambient electrical interference, which may obscure the essential neural signals. A comprehensive preprocessing pipeline was established to enhance the quality of EEG data for efficient feature extraction and classification.
Initially, all EEG recordings needed to be standardized to guarantee uniform signal duration among subjects. A bandpass filter with cutoff frequencies of 0.1 Hz to 70 Hz is used to reduce slow drifts and high-frequency noise, while preserving spectral regions critical for neurophysiological investigation. A notch filter at 50 Hz was utilized to eliminate power line interference. The filtering was performed using the FIR window method (firwin), preserving the signal's phase characteristics. 

We utilized a two-phase correction technique to mitigate non-neural artifacts. The preliminary stage was artifact subspace reconstruction (ASR) \cite{chang2020asr}, which minimizes transient artifacts by reconstructing the EEG signal using just the low-variance principal components during short overlapping intervals. This technique effectively reduces high-amplitude, non-stationary noise events, such as motion bursts and muscle contractions. In the second phase, Independent Component Analysis (ICA) \cite{makeig1995ica} was performed to extract statistically independent sources from the multichannel EEG data. Components identified as artifactual, particularly those associated with eye blinks, eye movements, or muscle noise, were eliminated \cite{arnaud2007artifacts}. The cleaned EEG signals were subsequently assembled by reassembling the remaining components.

Following artifact reduction, we retained 19 conventional EEG channels in accordance with the 10–20 international system \cite{george1999electrode}: \textit{Fp1, F3, C3, P3, O1, F7, T3, T5, Fz, Fp2, F4, C4, P4, O2, F8, T4, T6, Cz,} and \textit{Pz}. Each processed EEG recording was segmented into 5-second epochs using a sliding window with 50\% overlap, to maintain a balance between temporal accuracy and data quantity for model training. Ultimately, each segment underwent z-score normalization on a per-channel basis, ensuring that every channel displayed a mean of zero and a variance of one. This normalization facilitates fair comparison among channels and participants by eliminating amplitude scaling anomalies. The preprocessing pipeline from \textit{Dataset I} produced 6,440 artifact-free EEG segments from 56 people (28 with MDD and 28 HC). From \textit{Dataset II}, the identical process yielded 3,634 artifact-free EEG segments from 46 people (23 with MDD and 23 HC). The preprocessed datasets constitute the foundation for feature extraction and graph creation.

\subsection{Results on MDD Classification}

\begin{figure*}[htbp]
    \centering 
    \includegraphics[width=0.95\textwidth]{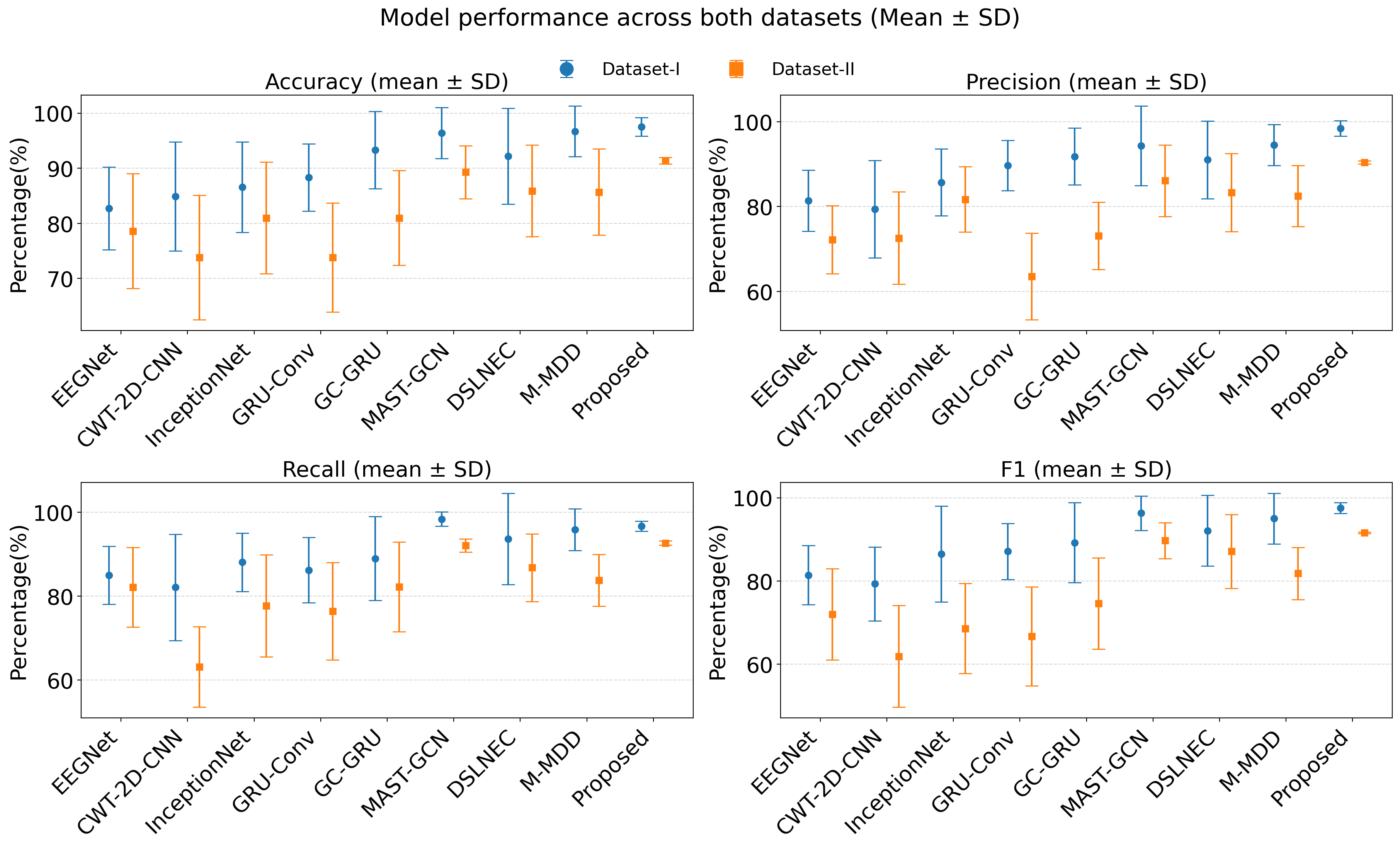}
    \caption{Evaluation of model performance on both datasets with error bars for all metrics.  Round markers represent Dataset-I and square markers represent Dataset-II. The error bars represent inter-fold variability. The plot shows the superior consistent performance of the proposed model in comparison to current baselines across both datasets.}
    \label{fig:mean_sd}
\end{figure*}

We evaluated the proposed framework against various CNN and RNN-based benchmarks, in addition to contemporary graph-based models. 
We provide a thorough comparison of our proposed method with existing methods in both datasets in Table~\ref{tab:performance_I}, Table~\ref{tab:performance_II}, and Figure~\ref{fig:mean_sd}.
The circular markers in Fig.~\ref{fig:mean_sd} represent \textit{Dataset-I}, whereas the square markers denote \textit{Dataset-II}, and the error bars indicate inter-fold variability.
The figure highlights two principal trends: (\textit{i}) ExPANet regularly attains the greatest mean values across all metrics, and (\textit{ii}) it displays the least variation, indicating robust generalization and stability among patients.

\paragraph{Dataset-I}
Table~\ref{tab:performance_I} indicates that ExPANet achieves an overall accuracy of 97.5\%, precision of 98.4\%, recall of 96.7\%, and F1-score of 97.3\%. 
In comparison to the leading previous graph model GC-GRU~\cite{luo2023gcgru} (accuracy 93.3\%, F1 89.2\%), this represents an absolute improvement of $+4.2$ percentage points in accuracy and $+8.1$ percentage points in F1.  
In comparison to convolutional baselines like EEGNet~\cite{lawhern2018eegnet} (accuracy 82.7\%) and InceptionNet~\cite{LEI2022103370} (accuracy 86.6\%), the enhancement surpasses $+10$ percentage points across all metrics, highlighting the advantages of graph-based modeling with feature-specific node properties.  
Statistical analysis indicates that these enhancements are significant with $p<0.01$ in comparison to EEGNet, CWT-2D-CNN, and GRU-Conv, and $p<0.05$ in relation to GC-GRU and M-MDD, as denoted by “**” and “*” in Table~\ref{tab:performance_I}.  
The small inter-fold standard deviation in Fig.~\ref{fig:mean_sd} further illustrates cross-subject dependability.

\newcommand{\pmark}[1]{\makebox[0pt][l]{$^{#1}$}}
\begin{table*}[htbp]
\centering
\caption{Subject-specific 10 Fold classification results on Dataset-I \cite{mumtaz2016dataset}}
\label{tab:performance_I}
\resizebox{0.9\linewidth}{!}{
\begin{tabular}{|l|c|c|c|c|c|}
\hline
Method & Reference & Accuracy (\%) & Precision (\%) & Recall (\%) & F1 Score (\%) \\
\hline
EEGNet \cite{lawhern2018eegnet} & J. Neural Eng. 2018 & 82.7\pmark{**}  & 81.4\pmark{**} & 85\pmark{**} & 81.4\pmark{**}  \\
CWT-2D-CNN \cite{SADATSHAHABI2021946} & Biocybernetics Biomed. Eng. 2021 & 84.9\pmark{**} & 79.4\pmark{**}  & 82.1\pmark{**}  & 79.3\pmark{**}   \\
InceptionNet \cite{LEI2022103370} & Biomed. Signal Process. Control 2022 & 86.6\pmark{*} & 85.7\pmark{*}  & 88.1\pmark{*}  &  86.5\pmark{*}   \\
{GRU-Conv} \cite{Xu2023emotion}  & Med. Biol. Eng. Comput. 2023 & 88.3\pmark{*} & 89.7\pmark{*}  & 86.2\pmark{*}  & 87.1\pmark{*} \\
GC-GRU \cite{luo2023gcgru} & IEEE TNSRE 2023 & 93.3\pmark{*} & 91.8\pmark{*} & 89\pmark{*} & 89.2\pmark{*}\\
MAST-GCN \cite{lumastgcn2024} & IEEE TAC 2024 & 96.4& 94.3& \textbf{98.4}& 96.3\\
DSLNEC \cite{Yuan2024latent} & IEEE TNNLS 2025 & 92.2\pmark{*} & 91\pmark{*} & 93.7\pmark{*} & 92.1\pmark{*}\\
M-MDD \cite{wang2025mmdd} & Neurocomputing 2025 & 96.7 &94.5 & 95.9 & 95 \\
Proposed &  - - & \textbf{97.5}  & \textbf{98.4}  & 96.7 &\textbf{97.3}  \\
\hline
\end{tabular}}
\smallskip

\parbox{0.85\linewidth}{
\footnotesize
Bold denotes the best experimental results in each metric. 
The performance of the comparative method, which is significantly lower than our method, is denoted as “*” ($p<0.05$) and “**” ($p<0.01$), as determined by a paired \textit{t}-test across cross-validation folds.
}
\end{table*}

\begin{table*}[htbp]
\centering
\caption{Subject-specific 10 Fold classification results on Dataset-II 
\cite{cavanagh2019multiple}}
\label{tab:performance_II}
\resizebox{0.9\linewidth}{!}{
\begin{tabular}{|l|c|c|c|c|c|}
\hline
Method & Reference & Accuracy (\%) & Precision (\%) & Recall (\%) & F1 Score (\%) \\
\hline
EEGNet \cite{lawhern2018eegnet} & J. Neural Eng. 2018 & 78.6\pmark{**}   & 72.2\pmark{**} & 82.1\pmark{**} &  72\pmark{**} \\
CWT-2D-CNN \cite{SADATSHAHABI2021946} & Biocybernetics Biomed. Eng. 2021 & 73.8\pmark{**} & 72.6\pmark{**}  & 63.1\pmark{**}  & 61.9\pmark{**} \\
InceptionNet \cite{LEI2022103370} & Biomed. Signal Process. Control 2022 &81\pmark{*} &81.7\pmark{*} & 77.7&\pmark{*}  68.6\pmark{**} \\
{GRU-Conv} \cite{Xu2023emotion}  & Med. Biol. Eng. Comput. 2023 & 73.8\pmark{**} & 63.5\pmark{**}  & 76.4\pmark{**}  &  66.7\pmark{**}\\
GC-GRU \cite{luo2023gcgru} & IEEE TNSRE 2023 & 81\pmark{*} & 73.1\pmark{*}& 82.2\pmark{*} & 74.6\pmark{*}\\
MAST-GCN \cite{lumastgcn2024} & IEEE TAC 2024 & 89.3 & 86.1& 92.1& 89.7\\
DSLNEC \cite{Yuan2024latent} & IEEE TNNLS 2025 & 85.9\pmark{*}& 83.3\pmark{*}& 86.8\pmark{*}& 87.1\pmark{*}\\
M-MDD \cite{wang2025mmdd} & Neurocomputing 2025 & 85.7\pmark{*} &82.5\pmark{*} & 83.8\pmark{*} & 81.8\pmark{*} \\
Proposed &  - - & \textbf{91.4}  &  \textbf{90.4} &\textbf{92.7}  & \textbf{91.6} \\
\hline
\end{tabular}}
\smallskip

\parbox{0.85\linewidth}{
\footnotesize
Bold denotes the best experimental results in each metric. 
The performance of the comparative method, which is significantly lower than our method, is denoted as “*” ($p<0.05$) and “**” ($p<0.01$), as determined by a paired \textit{t}-test across cross-validation folds.
}
\end{table*}

\paragraph{Dataset-II}
In the smaller and more heterogeneous cohort, ExPANet exhibits superior performance with an accuracy of 91.4\%, precision of 90.4\%, recall of 92.7\%, and F1-score of 91.6\% (Table~\ref{tab:performance_II}). 
The subsequent most effective method, M-MDD~\cite{wang2025mmdd}, achieves an accuracy of 85.7\% and an F1-score of 81.8\%, representing an absolute improvement of $+5.7$ percentage points and $+9.8$ percentage points, respectively.  
In comparison to the comprehensive graph model MAST-GCN~\cite{lumastgcn2024} (accuracy 89.3\%, F1 89.7\%), ExPANet demonstrates enhancements of $+2.1$ percentage points and $+1.9$ percentage points.  
All differences were confirmed to be statistically significant at $p<0.05$.  
The model's enhanced recall demonstrates its efficacy in accurately detecting MDD subjects, essential for clinical screening.
While its equivalent precision minimizes false positives among healthy controls, it assures balanced practicality.


The edge-gated attention module dynamically modifies the functional connectivity weights. 
The AxisMix block enhances the integration of features and dimensions. 
Collectively, these elements facilitate the model's development of stable, physiologically significant representations. 
The increase in both precision and recall indicates that the network prioritizes accurate diagnostic patterns rather than arbitrary connections.
The approach demonstrates stability across folds and datasets. 
ExPANet provides a dependable and comprehensible framework for the real-world EEG-based assessment of major depressive disorder.

\begin{figure*}[htbp]
  \centering
  \subfloat[Dataset-I: HC]{%
    \includegraphics[width=0.25\linewidth]{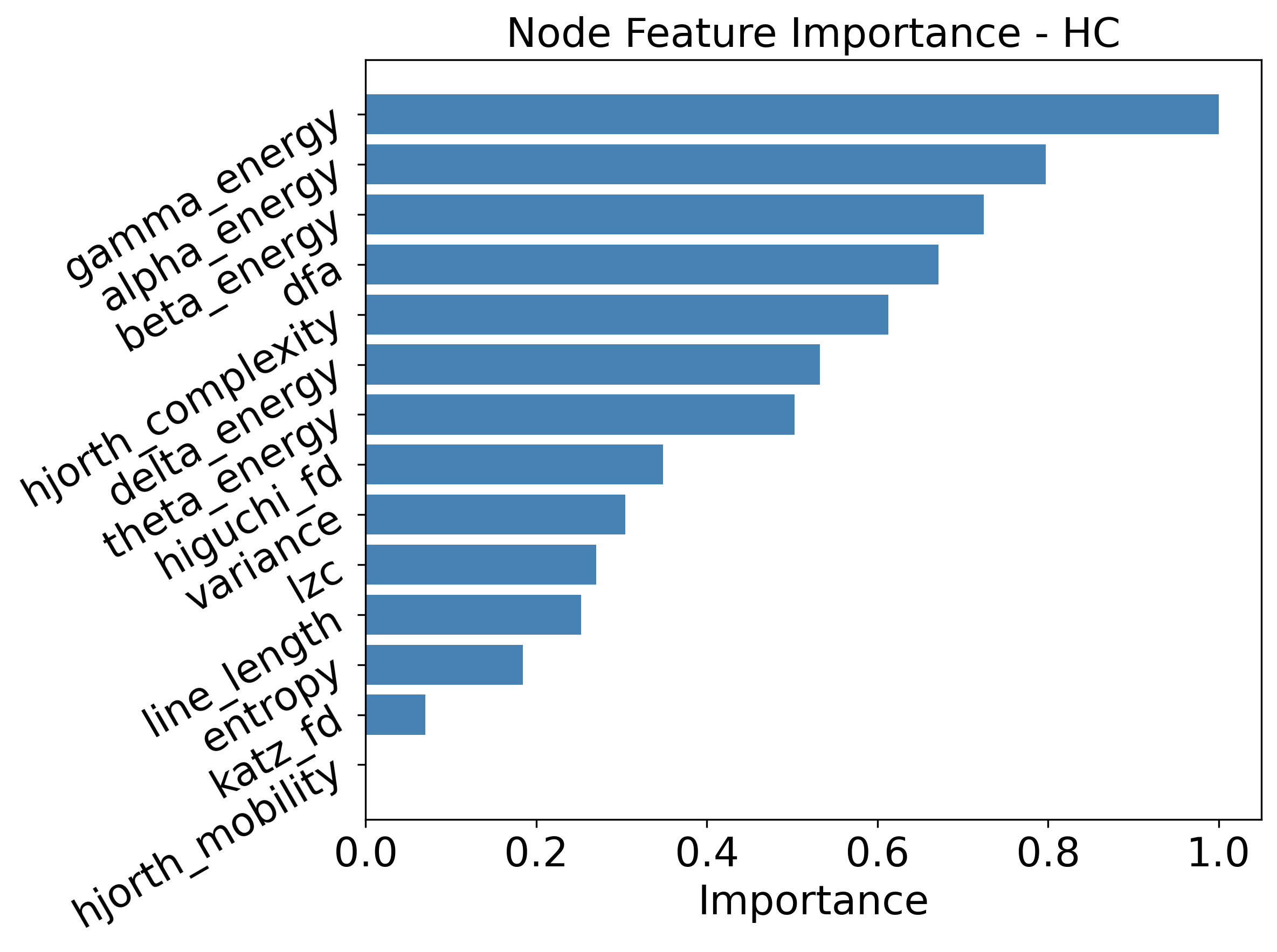}}
  \hfill
  \subfloat[Dataset-I: MDD]{%
    \includegraphics[width=0.25\linewidth]{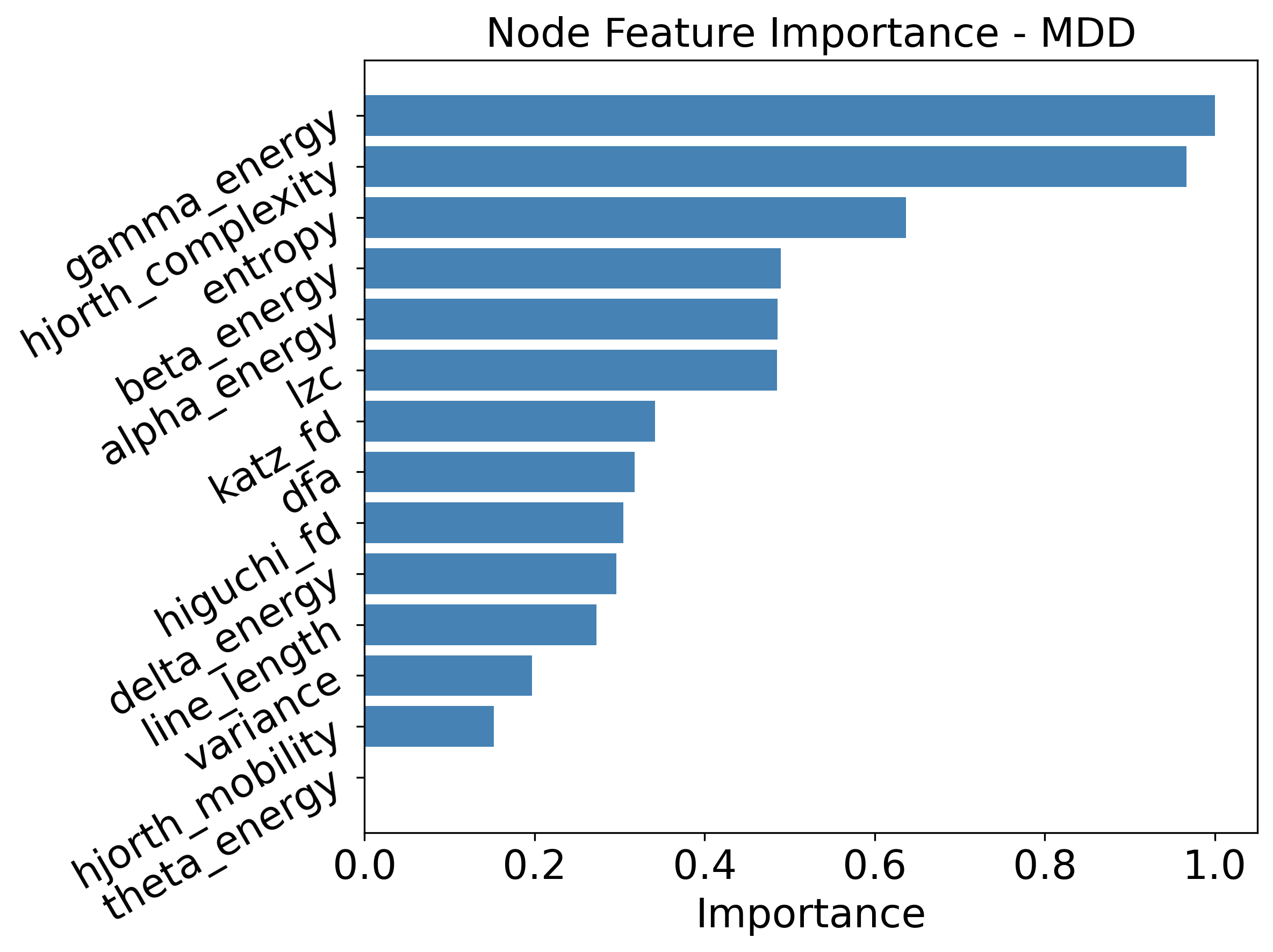}}
  \hfill
  \subfloat[Dataset-II: HC]{%
    \includegraphics[width=0.25\linewidth]{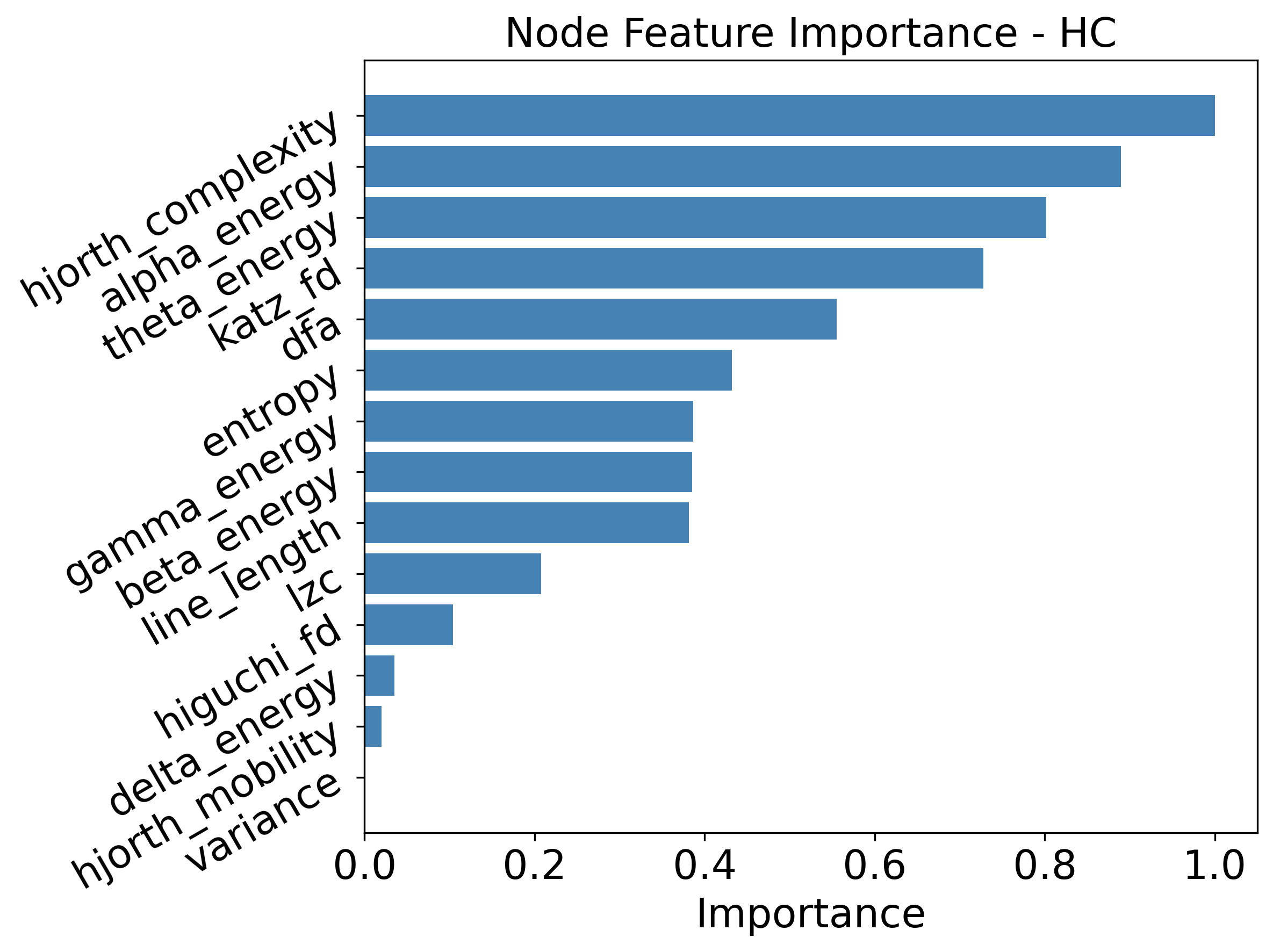}}
  \hfill
  \subfloat[Dataset-II: MDD]{%
    \includegraphics[width=0.25\linewidth]{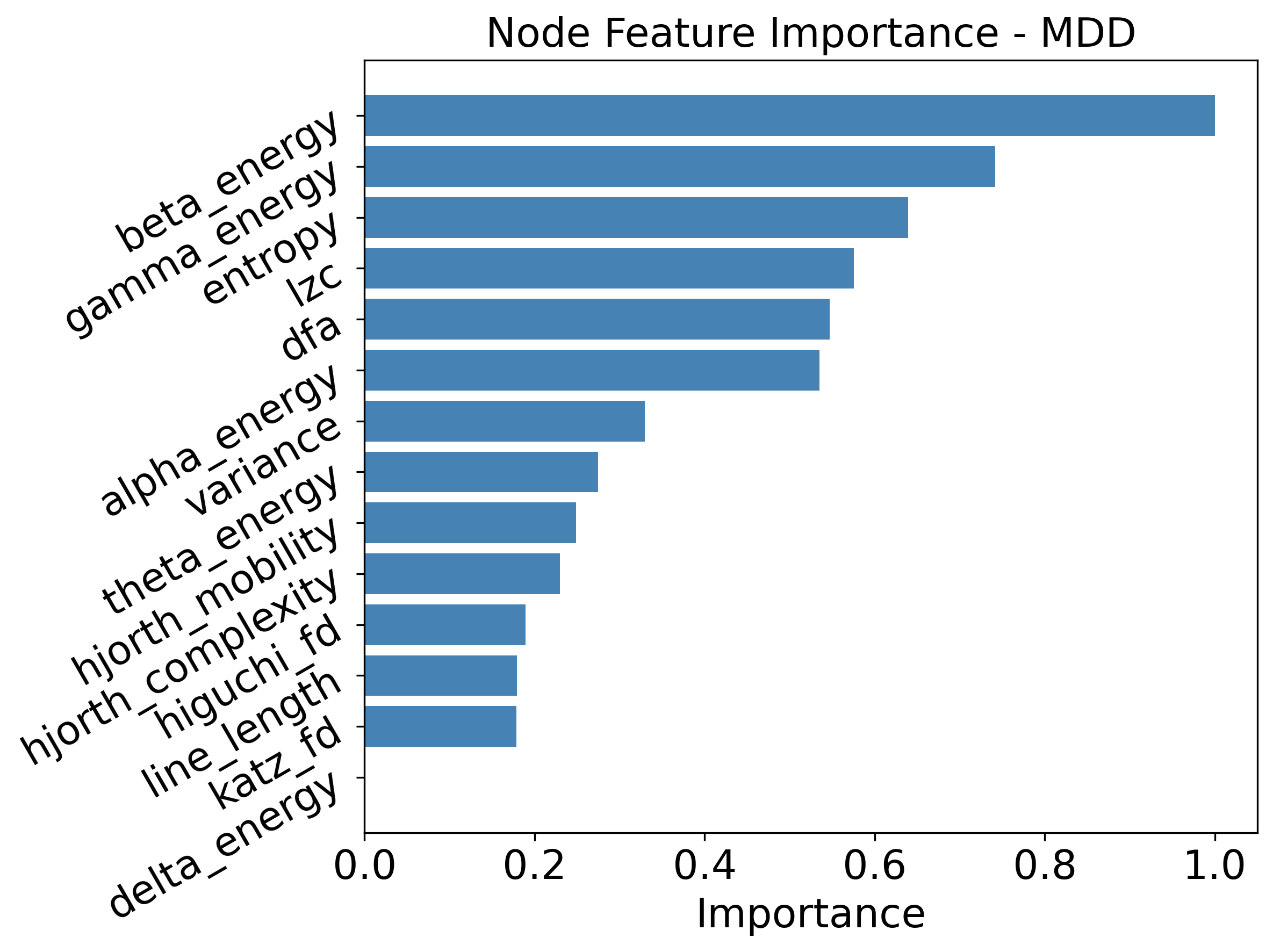}}
  \caption{Group-wise node-feature importance derived from GNNExplainer and averaged within HC vs MDD for each dataset. Bars are ordered low to high, to highlight the most discriminative features per group.}
  \label{fig:feat_importance}
\end{figure*}

\begin{figure*}[htbp]
  \centering
  \subfloat[Dataset-I: HC]{%
    \includegraphics[width=0.25\linewidth]{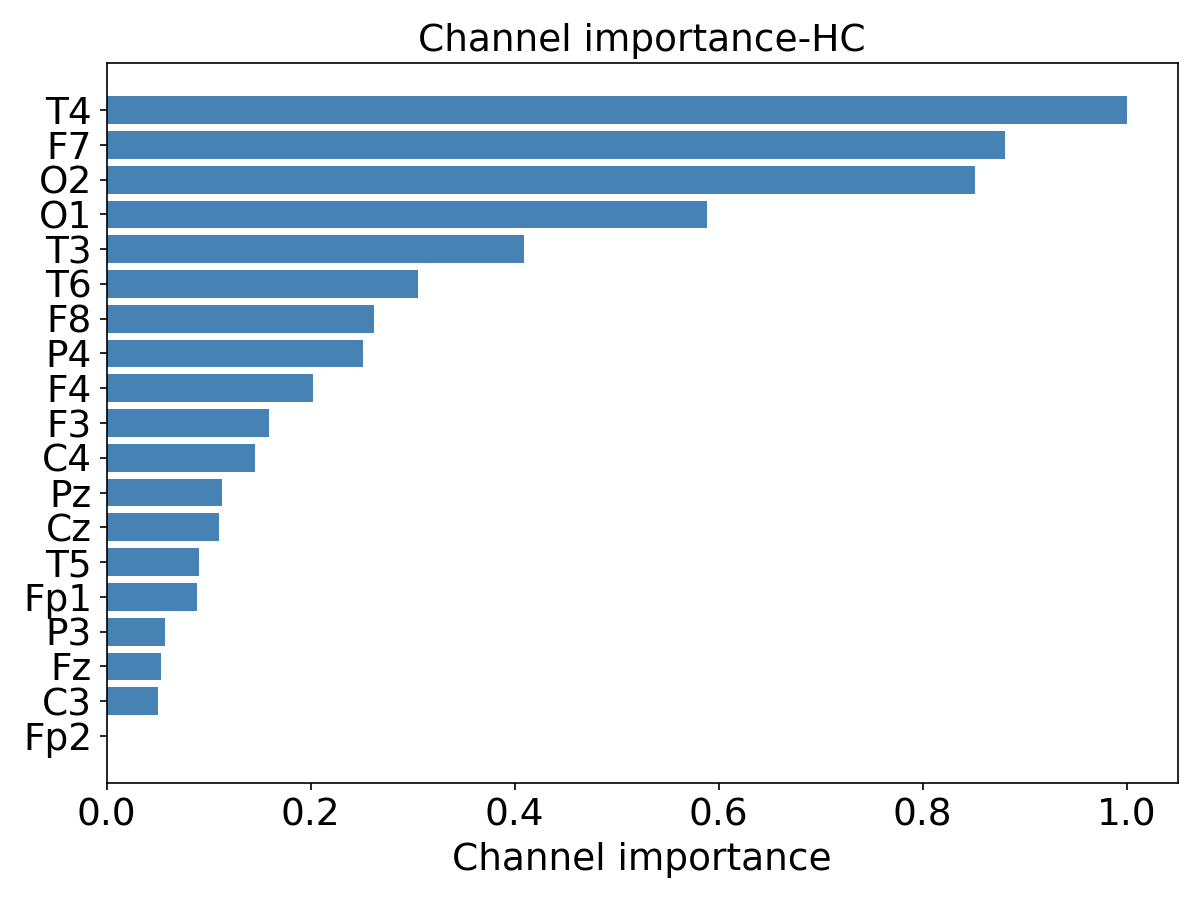}}
  \hfill
  \subfloat[Dataset-I: MDD]{%
    \includegraphics[width=0.25\linewidth]{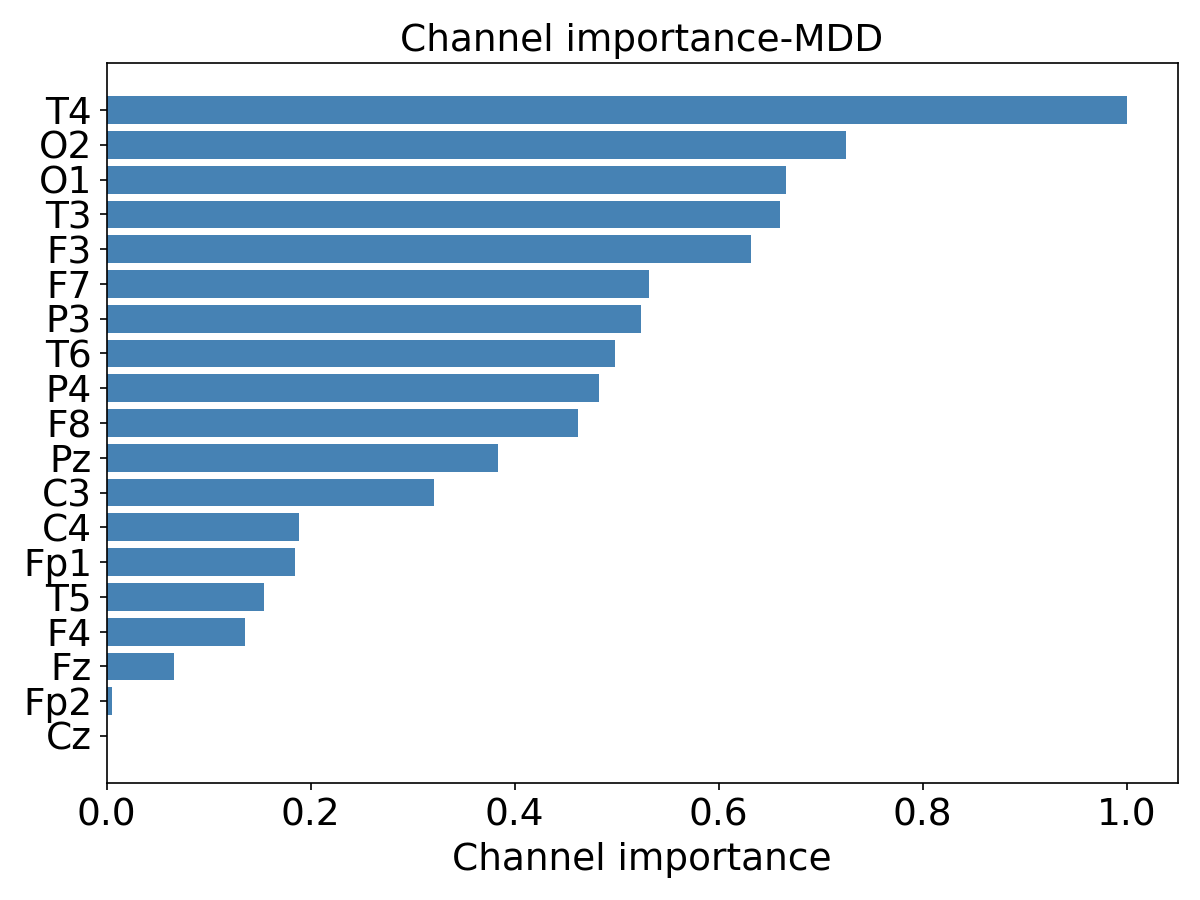}}
  \hfill
  \subfloat[Dataset-II: HC]{%
    \includegraphics[width=0.25\linewidth]{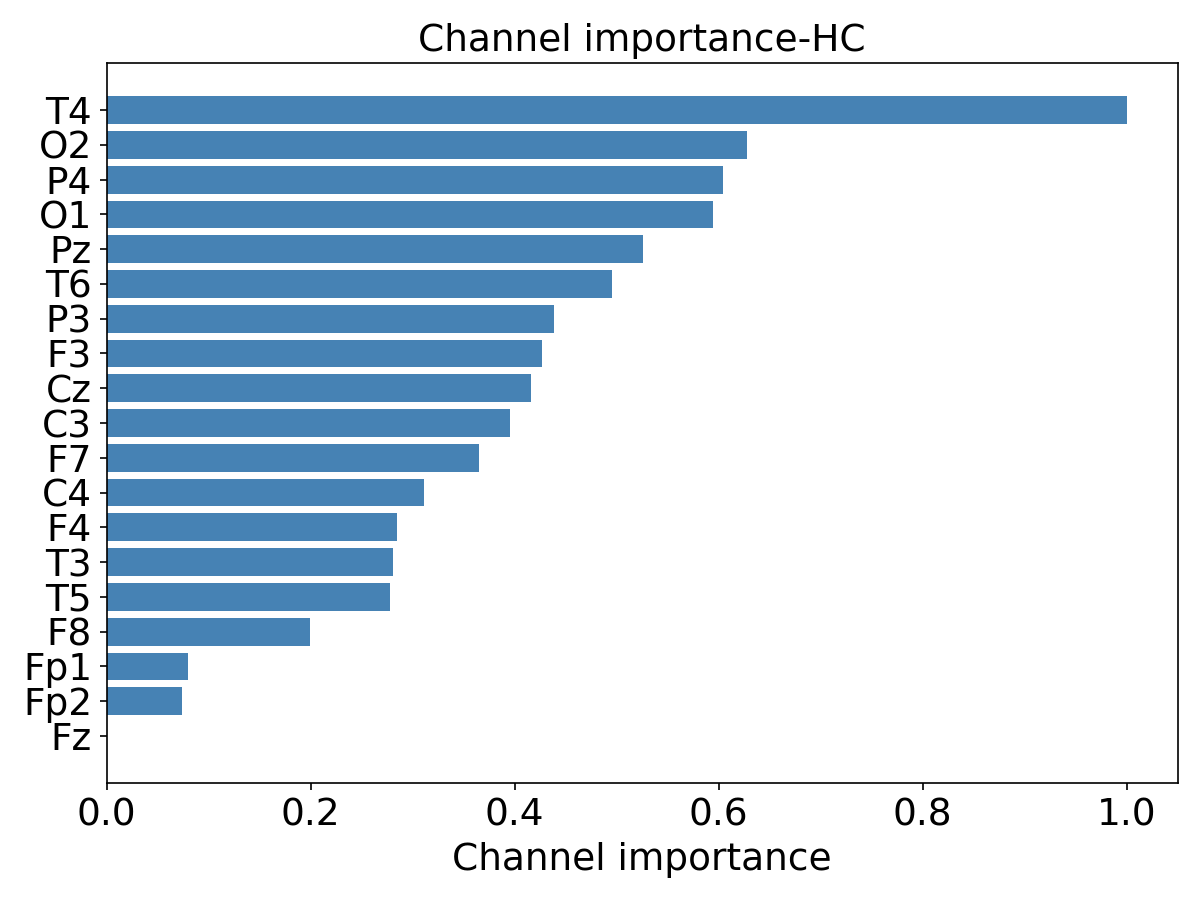}}
  \hfill
  \subfloat[Dataset-II: MDD]{%
    \includegraphics[width=0.25\linewidth]{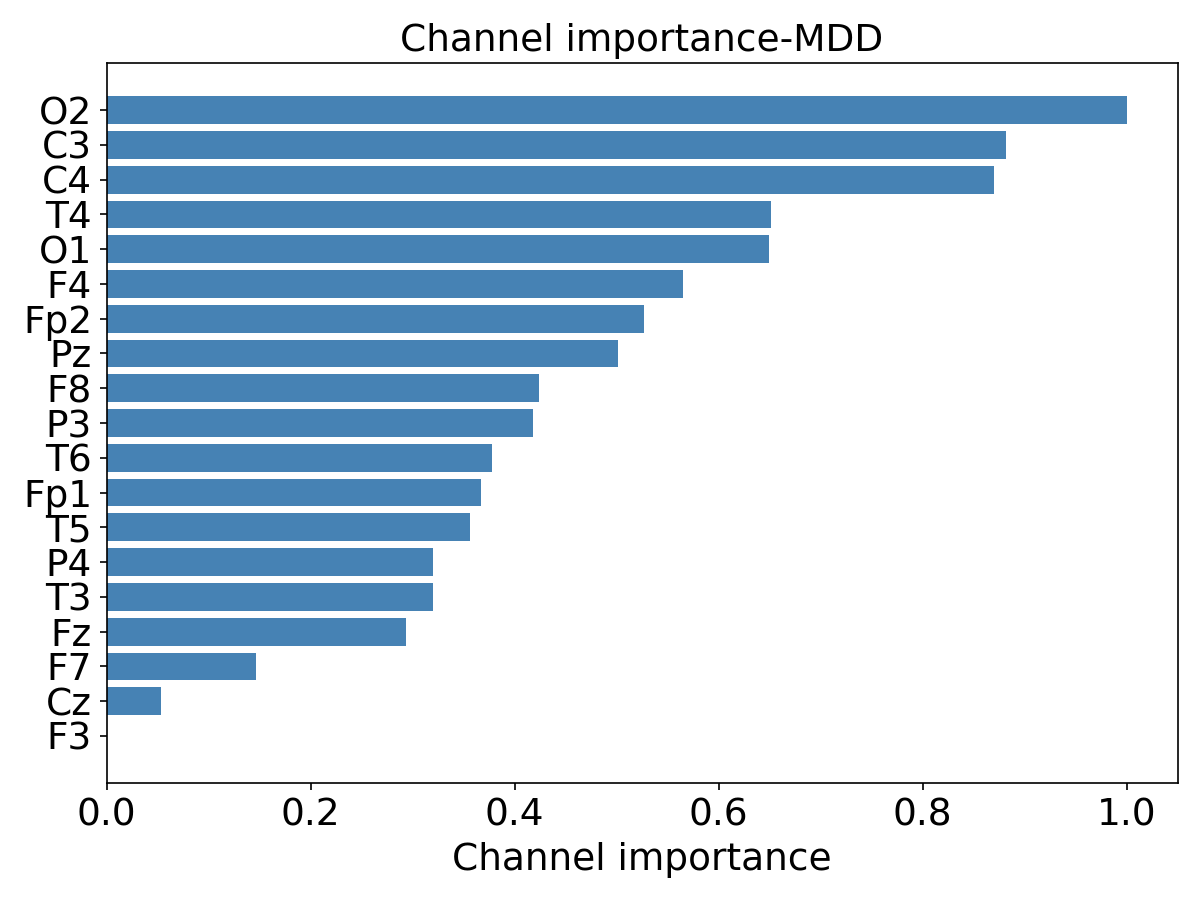}}
  \caption{Channel importance rankings for HC and MDD across both datasets. Fronto–central electrodes consistently rank higher than posterior sites.}
  \label{fig:chan_importance}
\end{figure*}

\subsection{Results on Explainability}

\subsubsection{Feature Importance}
We evaluated the relevance of node features by employing GNNExplainer masks aggregated at the group level (HC and MDD) for each dataset, as illustrated in Fig.~\ref{fig:feat_importance}. Per-group saliencies were normalized to a range of [0,1] and illustrated as ranking horizontal bar graphs.  Fundamental time-domain descriptors were consistently less significant than non-linear and long-range dependence metrics, validating previous findings \cite{vcukic2020successful}. Band-specific energies (alpha, beta, gamma) demonstrated dataset-dependent variations, indicating modified rhythmic dynamics in MDD, in alignment with the literature on frontal alpha asymmetry \cite{xie2023functional}.  These patterns offer a brief, feature-centric summary of the model's dependencies beyond simple EEG amplitudes.
\subsubsection{Channel Importance}
We analyzed group-averaged channel saliency as shown in Fig.~\ref{fig:chan_importance} utilizing the same aggregation method as for features. In both datasets, frontal and central electrodes (e.g., \textit{Fp1, Fz, Cz}) consistently exhibit higher rankings than posterior sites, suggesting that the model depends on fronto-central activity patterns for discriminating. This aligns with findings of modified frontal dynamics in MDD \cite{miljevic2023fc}, such as reduced frontal-theta variability and unusual synchronization, whereas posterior contributions are relatively weaker.
\subsubsection{Connectivity Importance}
\begin{figure}
  \centering
  \subfloat[Dataset-I]{%
    \includegraphics[width=0.48\linewidth]{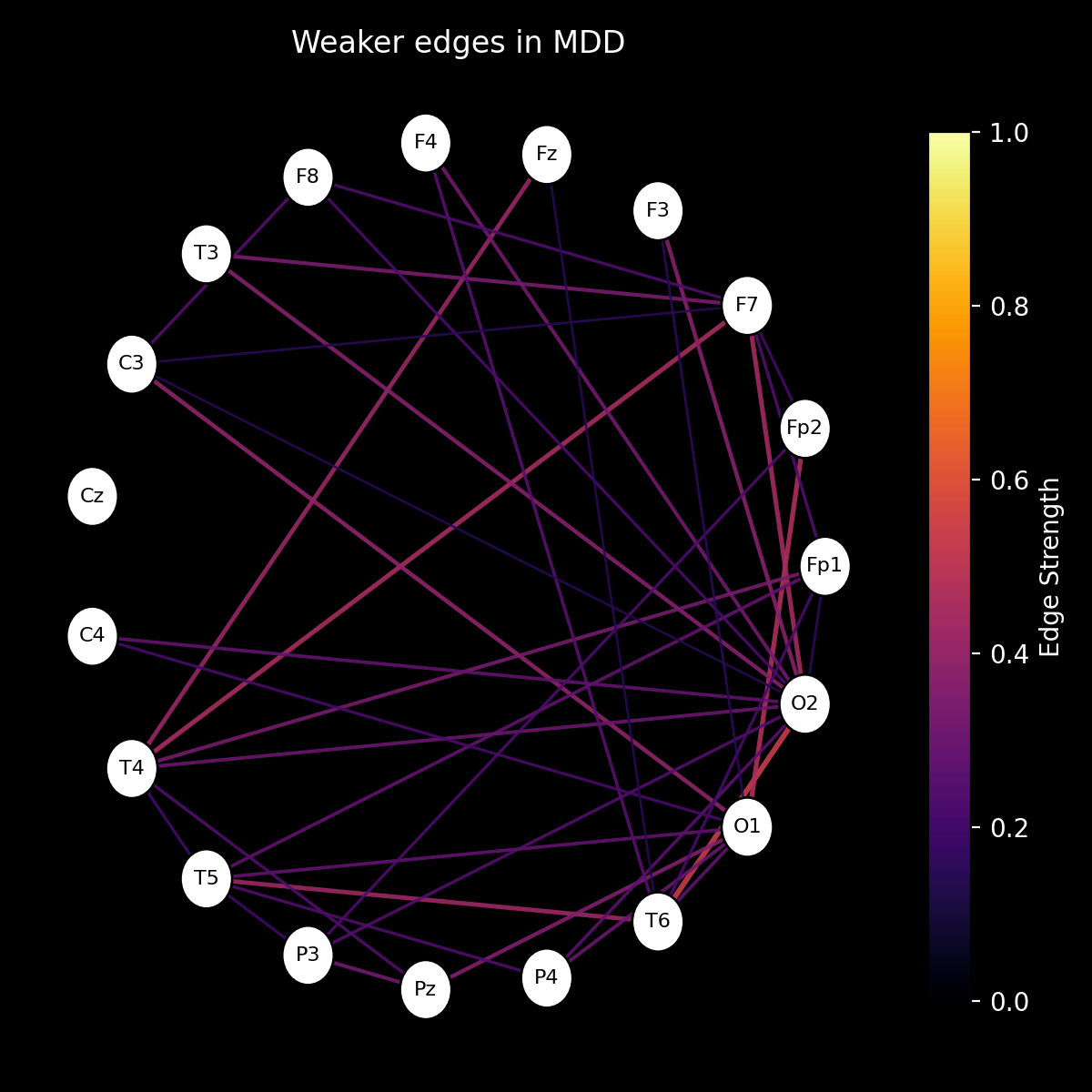}}
  \hfill
  \subfloat[Dataset-II]{%
    \includegraphics[width=0.48\linewidth]{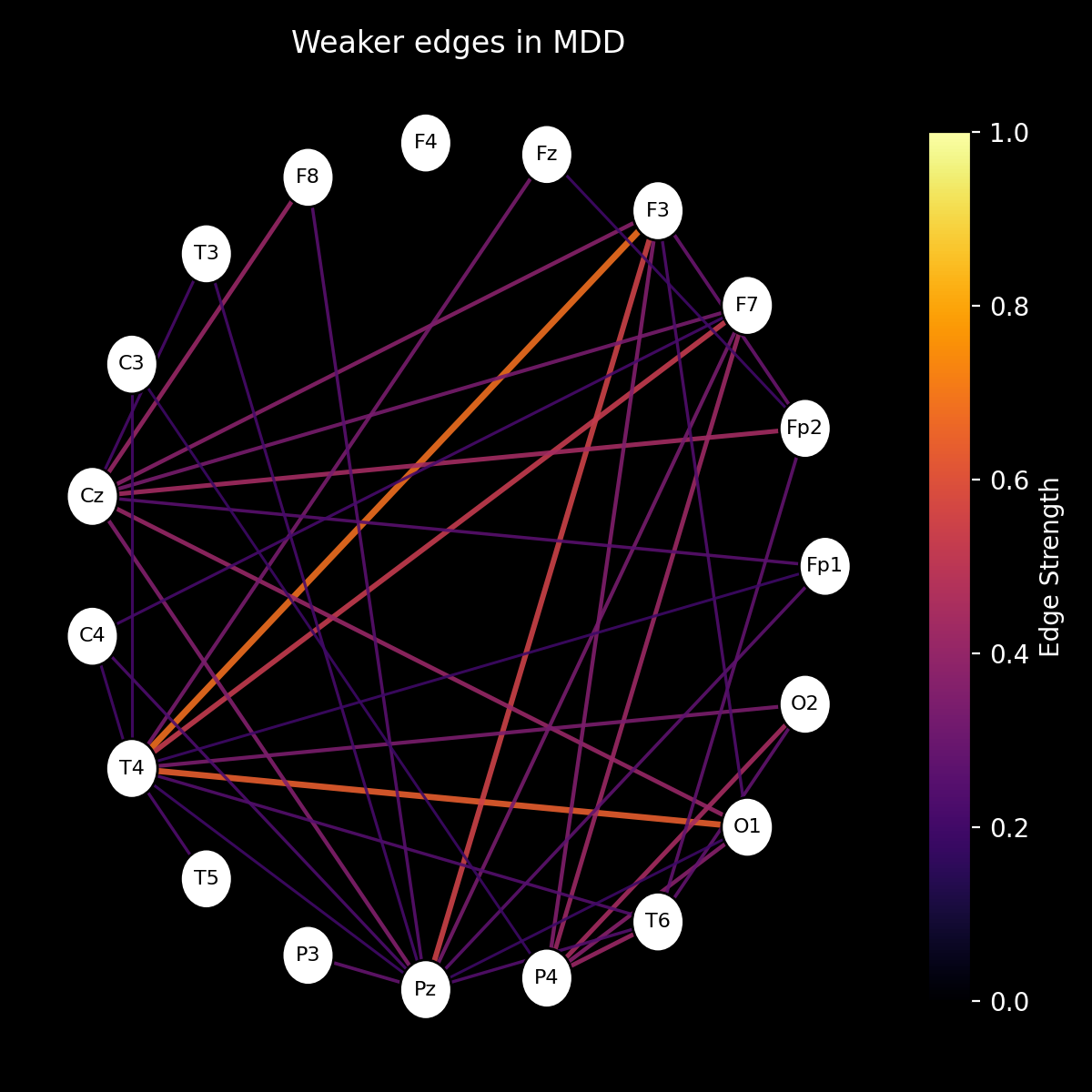}}
  \caption{Analysis of connectivity importance. Only edges consistently weaker in MDD than HC are shown, with a color scale indicating the degree of reduction (higher values indicate stronger weakening). Both datasets show weakened long-range occipito-temporal and occipito-frontal connections in MDD, although residual fronto-midline couplings remain dominant.}
  \label{fig:edge_importance}
\end{figure}

We conducted a further analysis of edge saliency across groups. Figure~\ref{fig:edge_importance} illustrates the subset of connections that were consistently diminished in MDD relative to HC. Healthy controls exhibited enhanced connectivity between occipital regions (\textit{O1/O2}) and temporal locations (\textit{T3/T4/T6}). They demonstrated strong linkages from the occipital to frontal regions (\textit{F3/F4/F7}) and occipito-parietal connectivity (e.g., \textit{O1-Pz, C3-O1}). Conversely, patients with MDD exhibited diminished long-range connectivity between the occipital-temporal and occipital-frontal regions. Additionally, they had comparatively enhanced fronto-midline and fronto-parietal couplings (e.g., \textit{Fp2–P4, Fp1–Pz, Fp2–Cz}). The results indicate that MDD is marked by deficient integration of sensory and cognitive processing pathways. The decrease in occipito-frontal and occipito-parietal connectivity signifies impaired information transfer, whereas the increased frontal-midline connections may represent compensatory processes. These findings align with prior EEG studies \cite{Leuchter2012qEEG, wang2024alterations} that indicated diminished posterior/occipital integration and abnormal enhancements in frontal connectivity in MDD.
\subsubsection{Intrinsic Interpretability}
\begin{figure}
  \centering
  \subfloat[Layer 1: attention]{%
    \includegraphics[width=0.48\linewidth]{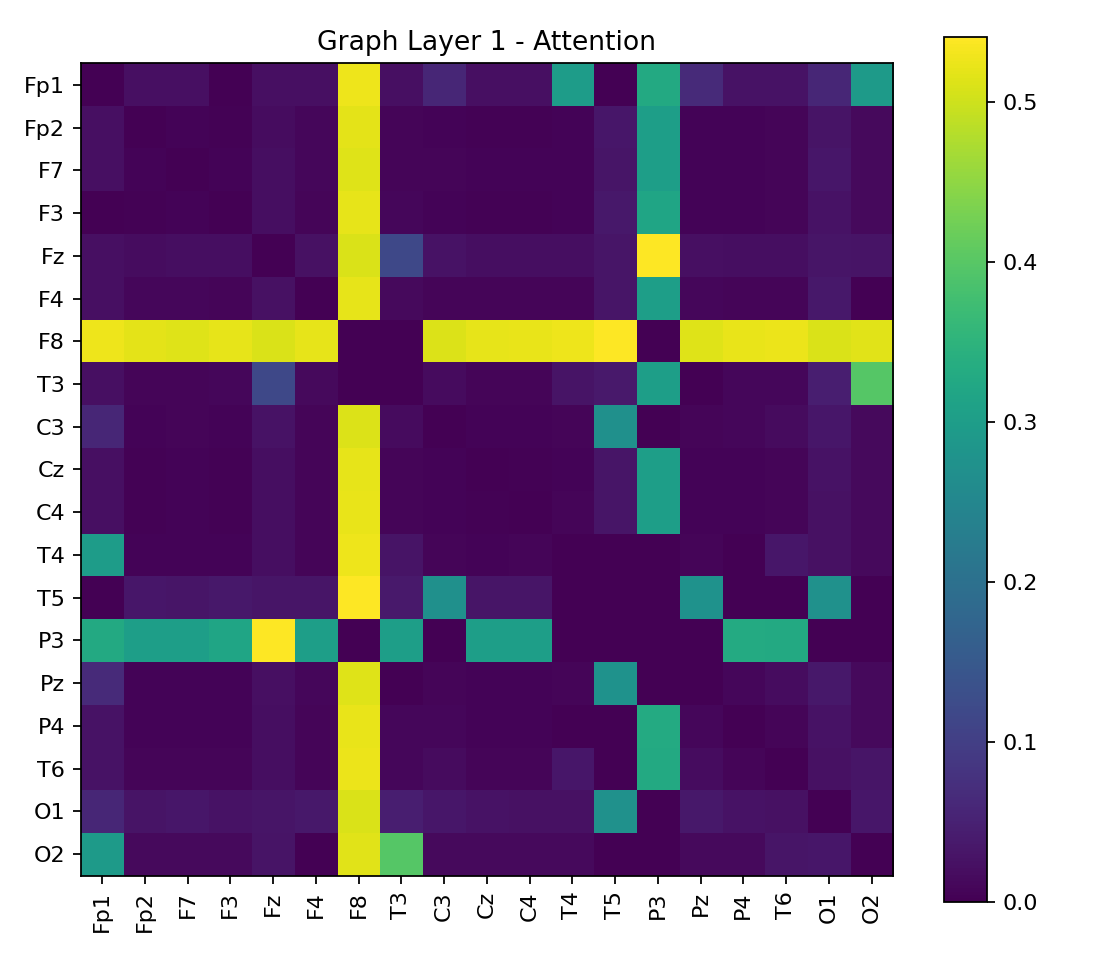}}
  \hfill
  \subfloat[Layer 2: attention]{%
    \includegraphics[width=0.48\linewidth]{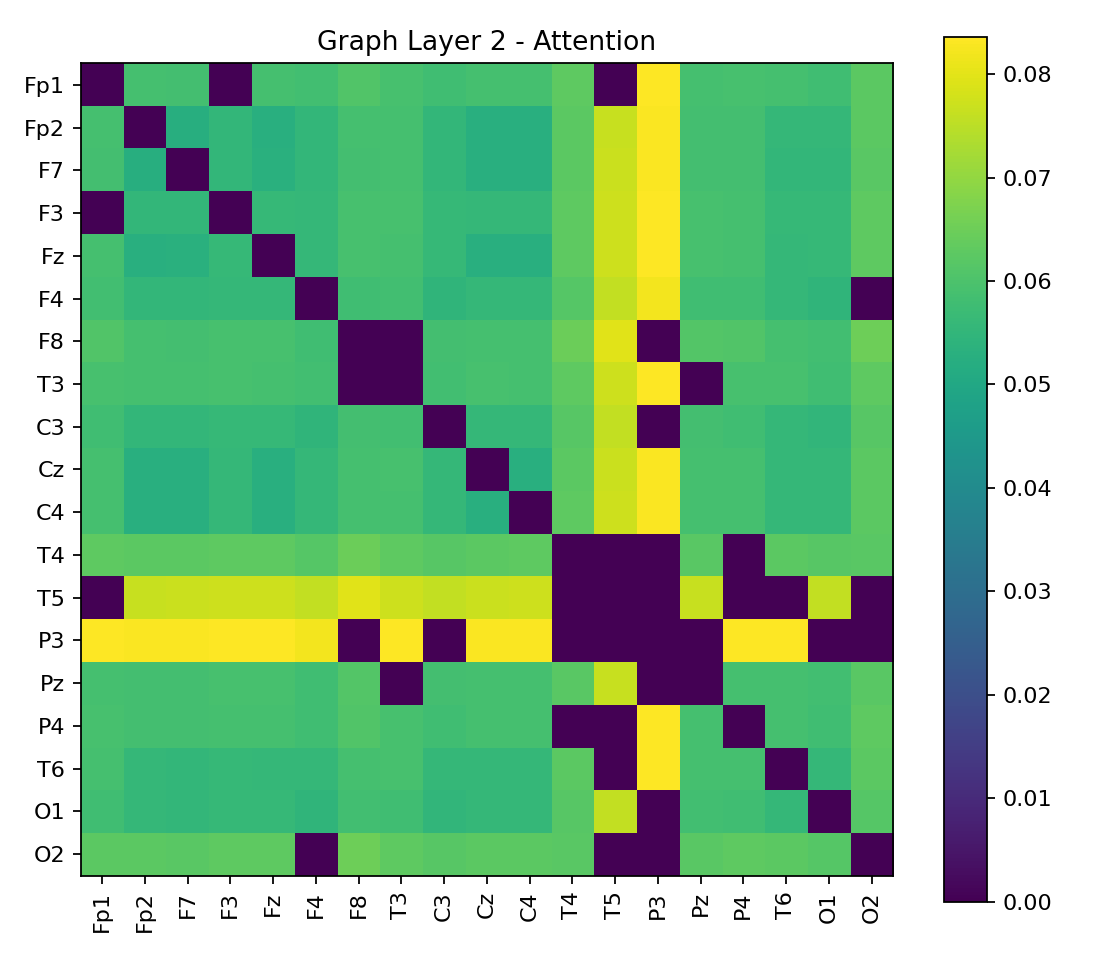}}
  \caption{Layer-wise attention heatmaps (mean $\alpha_{ij}$). 
  Higher values indicate greater importance assigned by the model to information flowing from channel $j$ to channel $i$. 
  The color scale reflects attention magnitude.}
  \label{fig:attn_heat}
\end{figure}

We illustrated layer-wise attention to demonstrate the channel–channel relationships that the model emphasizes during message passing (refer to \eqref{eq:edgeattention}). Figure~\ref{fig:attn_heat} displays the averaged attention matrix for the initial two layers of our proposed method. Each cell in the heatmap $(i,j)$ represents the normalized attention coefficient $\alpha_{ij}$, averaged across participants within the evaluation folds. Layer~1 focuses attention on local neighborhoods. Layer~2 allocates comparatively wider attention, addressing long-range relationships. These patterns corroborate the edge-saliency findings: channels associated with fronto-parietal and midline interactions generally get greater attention, whereas connections identified as weaker in MDD exhibit lower average attention. The post-hoc and intrinsic perspectives collectively offer a coherent understanding of the model's integration of spatial information.

\subsubsection{Group-level Attention Differences}
\begin{figure}
  \centering
  \subfloat[Layer 1: HC-MDD]{%
    \includegraphics[width=0.48\linewidth]{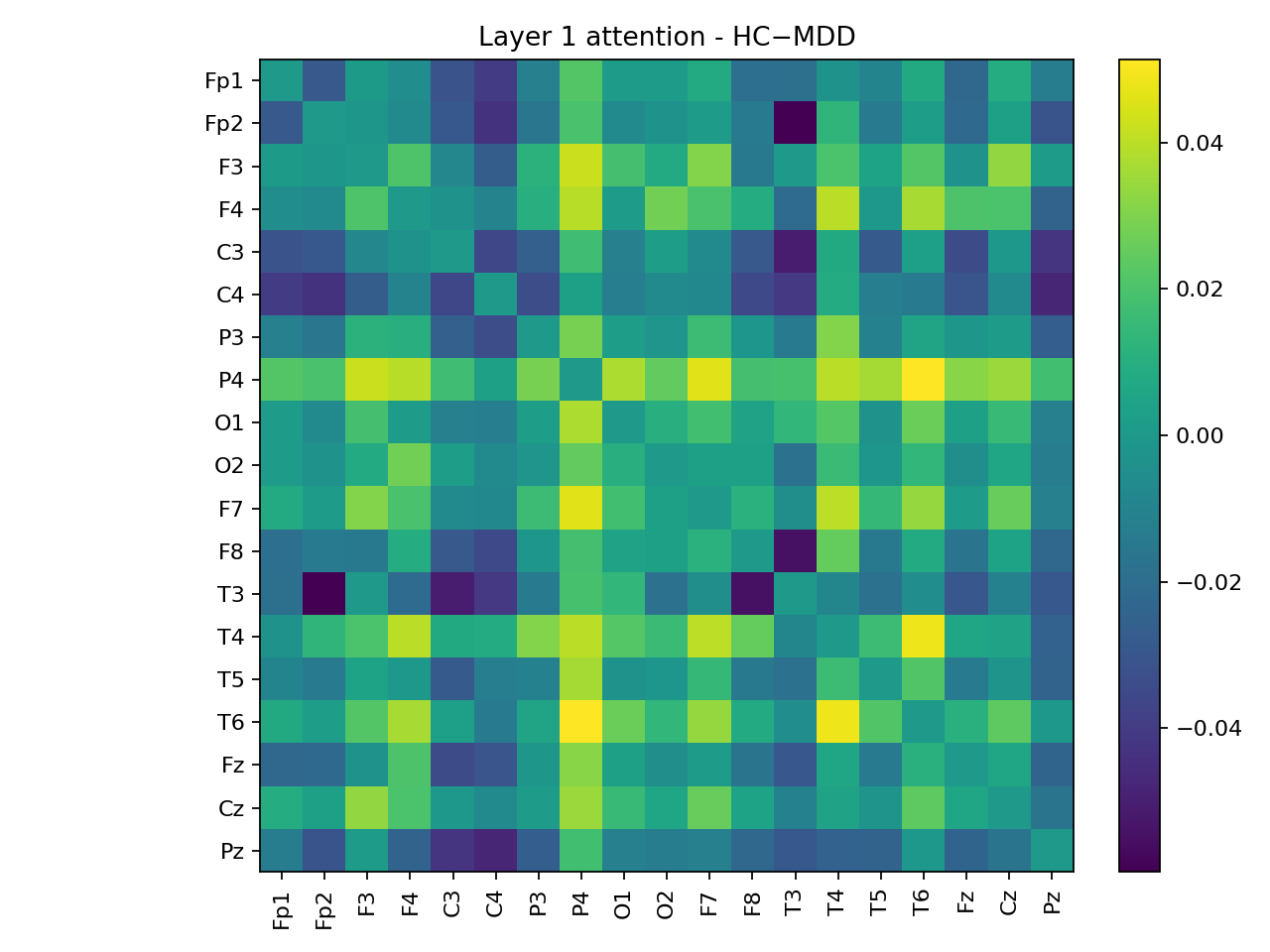}}
  \hfill
  \subfloat[Layer 2: HC-MDD]{%
    \includegraphics[width=0.48\linewidth]{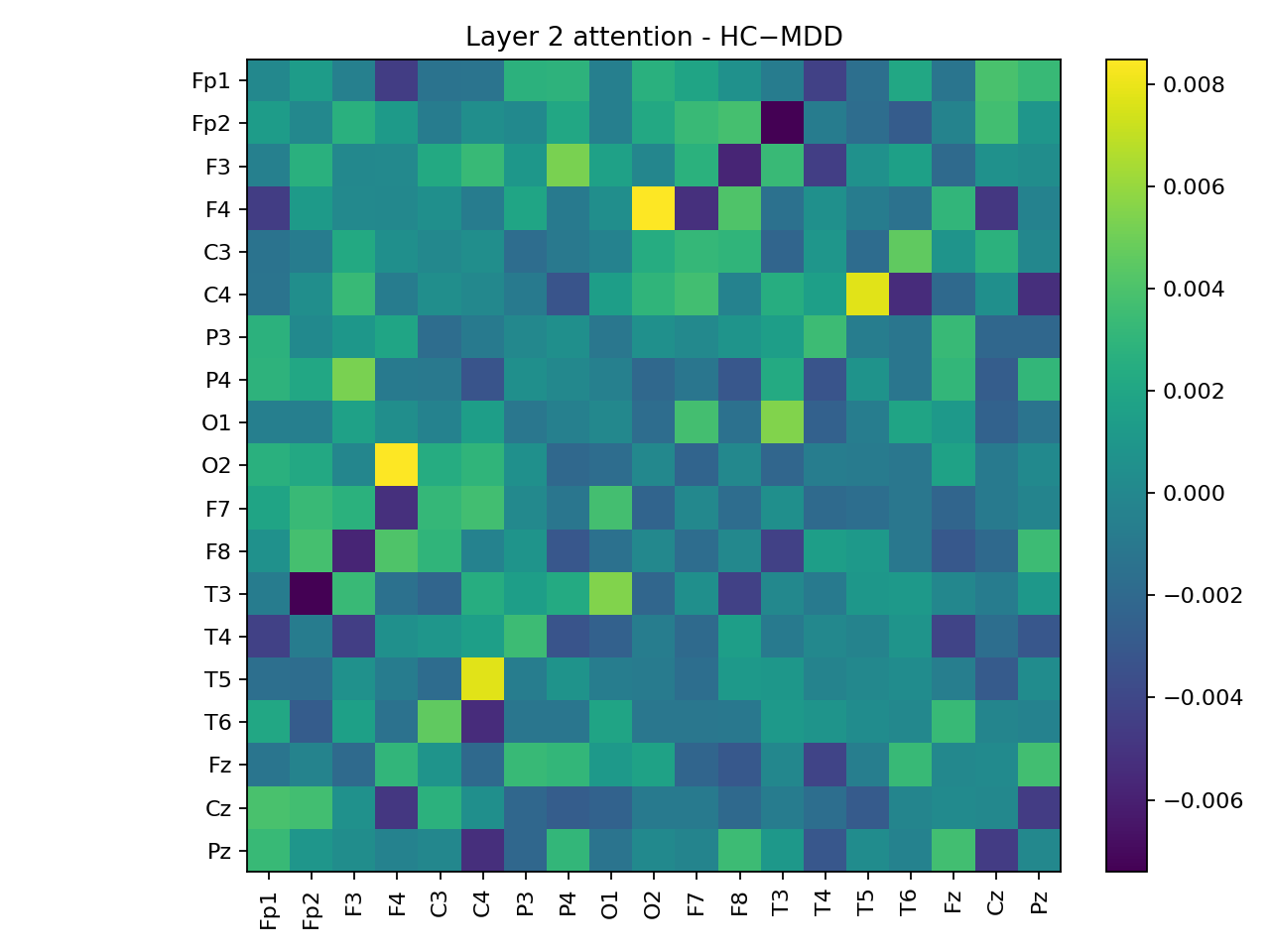}}
  \caption{Attention difference heatmaps (HC $-$ MDD).
  The color scale encodes the direction and magnitude of attention shift. Attention on long-range occipito–frontal/occipito–parietal relations is reduced in MDD, while some fronto–midline relations gain attention.}
  \label{fig:attn_diff}
\end{figure}

To analyze the alterations in attention associated with diagnosis, we calculated group-level attention matrices for HC and MDD and illustrated their differences. Figure~\ref{fig:attn_diff} illustrates the group-level disparities in attention weights (HC ${-}$ MDD) for the initial two graph neural network layers. Positive values signify that HC allocates increased attention to a specific connection, while negative values denote heightened focus in MDD. Layer~1 underscores significant disparities,  HC emphasizes long-range connections between parietal and occipital nodes, namely occipito-parietal and fronto-parietal couplings. Conversely, MDD demonstrates comparatively greater attention in frontal and midline interactions. In Layer~2, these disparities become broader yet continue to indicate diminished posterior integration in MDD and a residual enhancement of frontal–midline attention. These observations support the edge-level analysis:  MDD is marked by diminished dependence on posterior-anterior integration and compensatory enhancements in frontal neural activity.
\section{Conclusion}
\label{sec:conc}
This work presented an explainable graph-based deep learning framework for classifying MDD using EEG data. We developed feature-enriched brain graphs by modeling electrodes as nodes and functional connections as edges based on PLV. A novel graph attention architecture named ExPANet was developed to capture both node-level features and global connectivity patterns. The framework demonstrated enhanced performance relative to baseline CNN, RNN, and GNN models across two separate datasets. The main contribution of this work is the incorporation of explainability across many levels. Feature-level study indicated that nonlinear and fractal features are essential for discrimination. Channel-level saliency consistently indicated frontal and central electrodes, whereas edge-level explanations revealed decreased occipito–frontal and occipito–temporal connections in MDD. Weight analysis revealed a significant frontal–midline focus and diminished posterior integration in individuals. These results support previous clinical evidence, hence increasing confidence in the model's interpretability. 

Although encouraging outcomes, several constraints remain. The datasets employed are of moderate size and confined to particular paradigms. Additional validation on bigger and more heterogeneous populations is necessary to confirm generalizability. Furthermore, integrating EEG with additional modalities like fMRI or clinical questionnaires may enhance diagnostic reliability. In conclusion, our system demonstrated that high classification accuracy and interpretability can be attained simultaneously. The suggested approach explains functional features, electrodes, and connectivity patterns, thereby advancing the development of dependable and clinically usable EEG-based screening instruments for depression.

\bibliographystyle{IEEEtran}
\bibliography{references}


\end{document}